\documentclass[pra,twocolumn,superscriptaddress,showpacs,amsmath,amstex,amssymb,citeautoscript]{revtex4-1}

\usepackage[T1]{fontenc}
\usepackage[utf8]{inputenc}
\usepackage{lipsum}
\usepackage{amsfonts}
\usepackage{latexsym}
\usepackage{amsmath}
\usepackage{amssymb}
\usepackage{bm,bbm}
\usepackage{braket}
\usepackage{xcolor}
\usepackage{pifont}
\newcommand{\cmark}{\ding{51}}%
\newcommand{\xmark}{\ding{55}}%
\usepackage[mathscr]{euscript}
\usepackage[shortlabels]{enumitem}
\usepackage[justification=justified]{subcaption}
\usepackage{graphicx}
\usepackage{lipsum}
\allowdisplaybreaks
\usepackage{float}
\usepackage{graphicx}
\usepackage{dsfont}
\usepackage{comment}
\usepackage[colorlinks=true]{hyperref}  
\hypersetup{
    bookmarks=true,         
    unicode=false,          
    pdftoolbar=true,        
    pdfmenubar=true,        
    pdffitwindow=false,     
    pdfstartview={FitH},    
    pdftitle={Tunable superconductivity and Moebius Fermi surface},    
    pdfauthor={Scammell and Scheurer},     
    pdfsubject={},   
    pdfcreator={},   
    pdfproducer={}, 
    pdfkeywords={} {} {}, 
    pdfnewwindow=true,      
    colorlinks=true,       
    linkcolor=blue, 
    citecolor=blue,        
    filecolor=magenta,      
    urlcolor=blue           
}

\usepackage{lipsum}
\newcommand{\equref}[1]{Eq.~(\ref{#1})}

\newcommand{\figref}[1]{Fig.~\ref{#1}}
\newcommand{\refcite}[1]{Ref.~\onlinecite{#1}} 
\newcommand{\refscite}[1]{Refs.~\onlinecite{#1}}

\newcommand{\tableref}[1]{Table~\ref{#1}}

\newcommand{\pdagger}{{\phantom{\dagger}}}

\renewcommand{\approx}{\simeq}

\renewcommand{\vec}[1]{\boldsymbol{#1}}

\definecolor{wrongultramarine}{rgb}{1,0.5,0}

\begin{document}
\title{Tunable superconductivity and M\"obius Fermi surfaces \\ in an inversion-symmetric twisted van der Waals heterostructure}

\author{Harley D.~Scammell}
\affiliation{School of Physics, the University of New South Wales, Sydney, NSW, 2052, Australia}

\author{Mathias S.~Scheurer}
\affiliation{Institute for Theoretical Physics, University of Innsbruck, Innsbruck A-6020, Austria}

\begin{abstract}
We study theoretically a moir\'e superlattice geometry consisting of mirror-symmetric twisted trilayer graphene surrounded by identical transition metal dichalcogenide layers. We show that this setup allows to switch on/off and control the spin-orbit splitting of the Fermi surfaces via application of a perpendicular displacement field $D_0$, and explore two manifestations of this control: first, we compute the evolution of superconducting pairing with $D_0$; this features a complex admixture of singlet and triplet pairing and, depending on the pairing state in the parent trilayer system, phase transitions between competing superconducting phases. Second, we reveal that, with application of $D_0$, the spin-orbit-induced spin textures exhibit vortices which lead to ``M\"obius fermi surfaces'' in the interior of the Brillouin zone: diabatic electron trajectories, which are predicted to dominate quantum oscillation experiments, require encircling the $\Gamma$ point twice, making their M\"obius nature directly observable. We further show that the superconducting order parameter inherits the unconventional, M\"obius spin textures. Our findings suggest that this system provides a promising experimental avenue for studying systematically the impact of spin-orbit coupling on the multitude of topological and correlated phases in near-magic-angle twisted trilayer graphene.
\end{abstract}

\maketitle

\begin{figure}[b]
   \centering
    \includegraphics[width=0.9\linewidth]{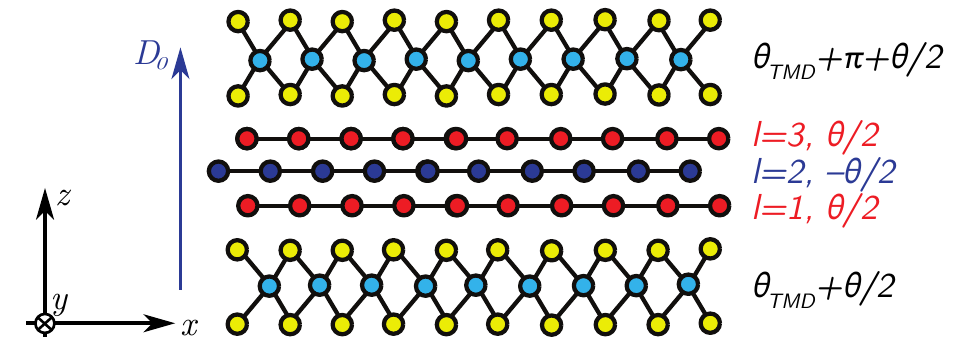}
    \caption{Moir\'e superlattice studied in this work, which consists of three layers of graphene ($l=1,2,3$) with alternating twist angle $\pm \theta/2$ symmetrically surrounded by TMDs with relative twist angle $\theta_{\text{TMD}} \approx 30^\circ$. A displacement field $D_0$ breaks inversion symmetry and, thus, allows to tune the spin-splitting of the bands.}
    \label{fig:overview}
\end{figure} 

Graphene moir\'e superlattices \cite{macdonald2019bilayer,andrei2020graphene} in the small twist angle regime, $\theta \approx 1^\circ - 2^\circ$, have emerged in the past few years as a versatile platform for realizing and probing a variety of correlated quantum many-body phases \cite{Cao2018_correlated, Cao2018_superconductivity,Yankowitz2019_TBG,Lu2019_TBG,Choi2019, Kerelsky2019, Xie2019, Jiang2019,Sharpe2019, Serlin2020,doi:10.1126/science.abc2836,FCIExp,MoireNematic,TMDNadjPerge,TMDNadjPerge_2,doi:10.1126/science.abh2889,Park2021_tTLG, Hao_2021,PauliLimitViolation,2021arXiv210912127K,2021arXiv210912631T,JiaTrilayer,2021arXiv211207127S,Jiang-Xiazi_diode,2021arXiv211001067D,2022NatPh..18..633J,2022arXiv220608354M}.  
While the intrinsic spin-orbit coupling (SOC) of graphene is very small \cite{PhysRevB.74.165310,PhysRevB.74.155426}, increasing it is expected to enrich the phenomenology of these systems even further: SOC opens up completely new avenues for stabilizing topological phases \cite{RevModPhys.82.3045,ZaletelSOC}, is expected to affect the energetic balance of closely competing \cite{PhysRevLett.122.246401,XieMacDonald2020,PhysRevX.10.031034,PhysRevB.103.205414,OurPRXTTLG,2022arXiv220405317W} instabilities as well as their form \cite{DiodeTheory}, and has an enormous potential for spintronics applications \cite{ReviewvdWs,AnotherReviewTwistronics,PhysRevB.106.L081406}.

Bringing graphene in close proximity to a transition metal dichalcogenide (TMD) layer, which involves heavier atoms, is known to induce SOC \cite{IndSOCExp1,IndSOCExp2,IndSOCExp3}; the form of the resultant SOC terms is well established for single-layer \cite{Gmitra2015,PhysRevB.99.075438,PhysRevB.100.085412,PhysRevB.104.195156,2022arXiv220609478L,PhysRevResearch.4.L022049}, and non-twisted multi-layer graphene \cite{PhysRevB.104.075126,PhysRevB.105.115126}. Importantly, the choice of TMD and the twist angle $\theta_{\text{TMD}}$ relative to the proximitzed graphene layer can be used to tune the strength and nature of the induced SOC \cite{PhysRevB.104.195156,PhysRevB.99.075438,PhysRevB.100.085412,2022arXiv220609478L,PhysRevResearch.4.L022049,PhysRevB.104.075126,PhysRevB.105.115126}. 
Although some experiments, see, e.g., \refscite{2021arXiv211207127S,doi:10.1126/science.abh2889}, clearly indicate that the proximitized TMD layers influence the correlated physics, in many cases it is not established what role the proximitized TMD layer plays for the observed phases. In general, understanding the impact of SOC on the correlated physics of graphene moir\'e systems is an open question.

To help elucidate the role of SOC, we here propose and analyze an inversion-symmetric graphene moir\'e superlattice setup, shown in \figref{fig:overview}, which has a key feature that the SOC-induced spin splitting of the Fermi surfaces can be tuned \textit{in situ}, by applying a perpendicular displacement field $D_0$. While the twisted graphene moir\'e system with the fewest number of layers---twisted bilayer graphene---exhibits spin-split bands already at $D_0=0$ \cite{ZaletelSOC}, the situation is different for mirror-symmetric twisted trilayer graphene (tTLG): it has inversion symmetry $I$, which persists when being surrounded symmetrically by TMD layers, see \figref{fig:overview}; together with time-reversal symmetry, $I$ guarantees pseudospin-degenerate bands, despite the presence of orbital SOC. Breaking it via $D_0 \neq 0$ allows to tune effective SOC terms lifting the bands' pseudospin degeneracy.  

In this Letter, we study the resulting tunability of the band structure and find that (for $\theta_{\text{TMD}}\approx 30^\circ$) the spin-orbit vector $\vec{g}_{\vec{k}}$ determining the spin-polarizations of the bands exhibits vortices in the interior of the moir\'e Brillouin zone.  
These imply that there is a filling fraction---found to be around half filling of the conduction or valence band---with Fermi surfaces that cross each other an odd number of times. We show that this feature can be observed in quantum oscillations where the dominant frequency corresponds to the \textit{sum} of the inner and outer spin-orbit-split Fermi sheets. To demonstrate the impact of the tunable SOC on the correlated physics of tTLG, we compute the superconducting order parameter as a function of $D_0$ assuming different dominant pairing states in the parent tTLG system. We see that $D_0$ not only allows to drive superconducting phase transitions and tune the order parameter, it also has the potential of being used as a tool to probe the nature of pairing in tTLG \cite{Park2021_tTLG, Hao_2021,PauliLimitViolation,2021arXiv210912127K,2021arXiv210912631T,JiaTrilayer,2021arXiv211207127S, Jiang-Xiazi_diode}, which is currently under debate.

\begin{figure}[tb]
   \centering
    \includegraphics[width=1.0\linewidth]{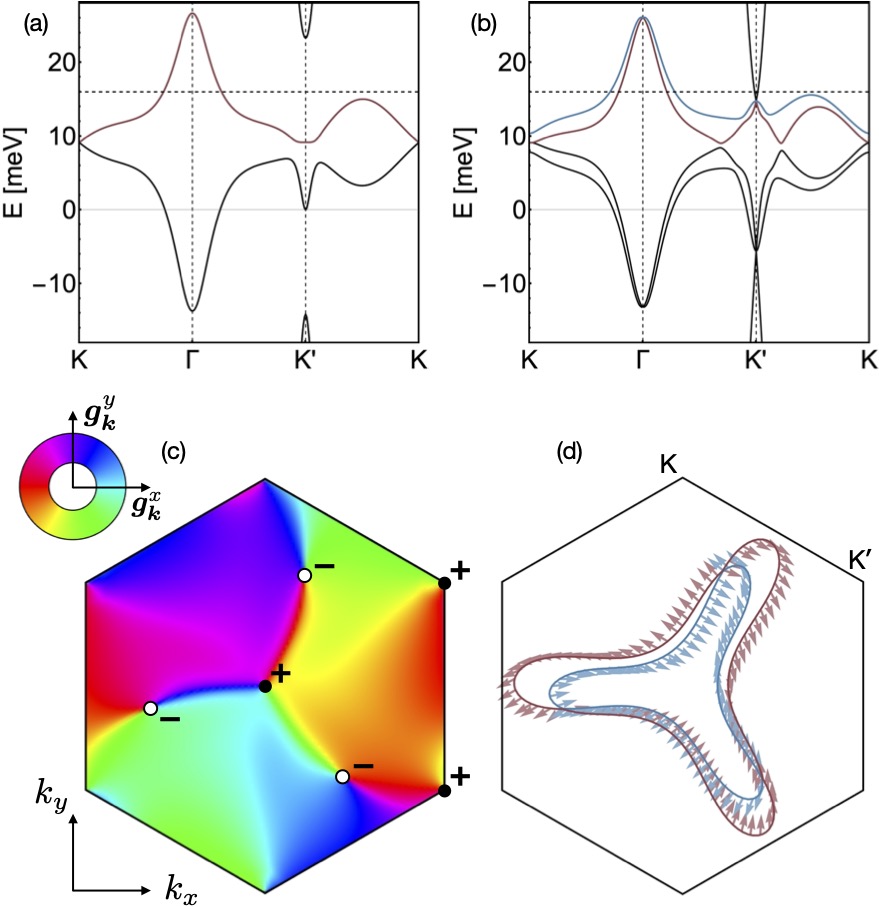}
    \caption{Band structure, including Fermi surfaces and spin texture in valley $\eta = +$. Here twist angle $\theta=1.75^\circ$, $\mu=\mu^*\approx16$ meV, and $\lambda_{\text{R}}= 20\,\textrm{meV}$. The cut through the band structure in (a) at $D_0=0$ shows spin-degenerate bands. The degeneracy is removed when $D_0\neq 0$ as can be seen in (b), where $D_0= 20\,\textrm{meV}$. Using the same parameters as in (b), the direction of the two non-zero components of $\bm g_{\bm k}$ in (c) reveals three vortices ($+$) and three anti-vortices ($-$). When $\mu=\mu^*$, we obtain vanishing Fermi-surface splitting, see (d), at their $C_{3z}^s$-related $\vec{k}$-space locations, leading to a M\"obius winding of the spin texture (arrows).
    }
    \label{fig:Bandstructure}
\end{figure}

\vspace{1em}\textit{Non-interacting bandstructure.}---To model the non-interacting bandstructure of the system in \figref{fig:overview} we use an appropriate generalization of the continuum-model description of twisted bilayer graphene \cite{dos2007graphene,bistritzer2011moire,dos2012continuum,KhalafKruchkov2019,2019arXiv190712338L,PhysRevB.103.195411}. The associated Hamiltonian is diagonal in the graphene layers' valley degree of freedom, $\eta=\pm$, and consists of four terms, $h_\eta = h_\eta^{(g)} + h_\eta^{(t)} + h^{(\text{SOC})}_\eta+ h^{(D)}_\eta $ \cite{Supplement}. Here $h_\eta^{(g)}(\vec{\nabla})$, $\vec{\nabla}=(\partial_x,\partial_y)$, and $h_\eta^{(t)}(\vec{r})$, $\vec{r}=(x,y)$, capture, respectively, the Dirac cones, rotated by $(-1)^l \theta/2$ and with opposite chirality for $\eta=\pm$, of the three layers, $l=1,2,3$, and the tunneling between them, respectively. The latter is spatially modulated on the emergent moir\'e scale and reconstructs the graphene cones into moir\'e bands. As a result of the mirror symmetry \cite{KhalafKruchkov2019}, $\sigma_h$, the bandstructure of $h_\eta^{(g)} + h_\eta^{(t)}$ is that of twisted bilayer graphene ($\sigma_h$-even) and single-layer graphene ($\sigma_h$-odd bands) in valley $\eta$. Surrounding tTLG with TMD layers as shown in \figref{fig:overview} induces SOC \cite{Gmitra2015,PhysRevB.99.075438,PhysRevB.100.085412,PhysRevB.104.195156,2022arXiv220609478L,PhysRevResearch.4.L022049} in the outer two layers as described by
\begin{equation*}
    h^{(\text{SOC})}_\eta = \sum_{l=1,3} (-1)^{\frac{(l-1)}{2}} P_l \left[ \lambda_{\text{I}} s_z \eta + \lambda_{\text{R}} \left(\eta \rho_x s_y - \rho_y s_x \right)\right],
\end{equation*}
where $P_l$ projects onto the $l$th graphene layer and $\rho_j$ ($s_j$), $j=x,y,z$, are Pauli matrices in sublattice (spin) space. The relative minus sign between the $l=1,3$ terms in $h^{(\text{SOC})}_\eta$ is dictated by the inversion symmetry, $I$, of the heterostructure. Note that $h^{(\text{SOC})}_\eta$ not only breaks spin-rotation symmetry, but also $\sigma_h$, which can be seen in the bandstructure shown \figref{fig:Bandstructure}(a): the graphene Dirac cone of the $\sigma_h$-odd sector, located around the $K'$ point for the valley shown, hybridizes with the $\sigma_h$-even, twisted-bilayer graphene sector. Despite the presence of SOC, all bands are still doubly degenerate, which follows from the momentum-space-local, anti-unitary symmetry $I \Theta_s$, with $\Theta_s$ being spin-$1/2$ time-reversal, that obeys $(I \Theta_s)^2 = -\mathbbm{1}$.

Applying a displacement field $D_0$---as is routinely done in experiments on tTLG \cite{Park2021_tTLG, Hao_2021,PauliLimitViolation,2021arXiv210912127K,2021arXiv210912631T,JiaTrilayer,2021arXiv211207127S, Jiang-Xiazi_diode} and captured by the last term, $h^{(D)}_\eta$, in the Hamiltonian---breaks $I$ and $I \Theta_s$. As such, the pseudospin-degeneracy of the bands at $D_0=0$ will be removed when $D_0\neq 0$, see \figref{fig:Bandstructure}(b), providing direct experimental control over the inversion-antisymmetric SOC terms, $\vec{g}_{\vec{k}} \neq 0$, in the effective Hamiltonian, $h^{\text{eff}}_{\vec{k},\eta} = s_0 \xi_{\eta\cdot\vec{k}} + \eta \,\vec{g}_{\eta\cdot\vec{k}}\cdot\vec{s}$, for the bands of tTLG near the Fermi level. To further discuss the form of $\vec{g}_{\vec{k}}$, let us focus on the limit $\theta_{\text{TMD}} = 30^\circ$, where $\lambda_{\text{I}}$ vanishes by symmetry \cite{PhysRevB.99.075438,PhysRevB.100.085412,PhysRevB.104.195156,2022arXiv220609478L} and $(\vec{g}_{\vec{k}})_z = 0$ as a consequence of $C_{2z}^s\Theta_s$ ($C_{nz}^s$ is the $n$-fold rotation symmetry along $z$). As can be seen in \figref{fig:Bandstructure}(c), the remaining two components of $\vec{g}_{\vec{k}}$ exhibit vortices at three, $C_{3z}^s$-related, generic positions $\vec{k}=\vec{k}^*_j$, $j=1,2,3$, in the Brillouin zone, compensating those with opposite chirality at $\Gamma$, $K$ and $K'$. Since these vortices---and, thus, the associated zeros of $|\vec{g}_{\vec{k}}|$ and type-II Dirac cones in the band structure---cannot be adiabatically removed, we are guaranteed to find a chemical potential $\mu=\mu^*$ where the Fermi surface of $\xi_{\vec{k}}$ 
crosses all three $\vec{k}^*_j$. For $\mu=\mu^*$ ($\mu\approx \mu^*$), the spin-orbit splitting of the Fermi surfaces of the system has to (almost) vanish at these three points, see \figref{fig:Bandstructure}(d).  
Having an odd number of points on the Fermi surface where the spin splitting (almost) vanishes also leads to exotic spin textures: following the spin polarizations of the Bloch states [also shown in \figref{fig:Bandstructure}(d)] diabatically along the Fermi surface, one ends up on the other Fermi sheet after one full revolution, akin to an object traversing a M\"obius strip. For this reason and due to the related, but not identical concept of ``M\"obius fermions'' \cite{ReviewMoebiusFSs}, which occur as edge states crossing the zone boundary in systems with non-symmorphic symmetries, we refer to the Fermi surfaces at $\mu=\mu^*$ as ``M\"obius Fermi surfaces''.
Note that M\"obius Fermi surfaces are only possible since a single valley neither has two-fold out-of-plane rotation nor time-reversal symmetry, which would necessarily lead to an even number of crossing points.  

\vspace{1em}\textit{Quantum oscillations.}---As our first example of an observable phenomenon associated with these M\"obius Fermi surfaces, we discuss quantum oscillations, which are routinely observed in resistivity in small-angle graphene moir\'e systems \cite{Cao2018_correlated,PabloQOs,QO_TDBG}. In the semi-classical picture, the conduction electrons will undergo periodic orbits on quantized constant-energy contours in momentum space when a perpendicular magnetic field $B\neq 0$ is applied. The oscillations of physical quantities are associated with these contours crossing the Fermi level and the frequency $F$ of the oscillations as a function of $1/B$ are proportional to the momentum-space area enclosed by the respective Fermi-surface contours \cite{Onsager,lifshitz1956theory}. 

\begin{figure}[t]
   \centering
    \includegraphics[width=\linewidth]{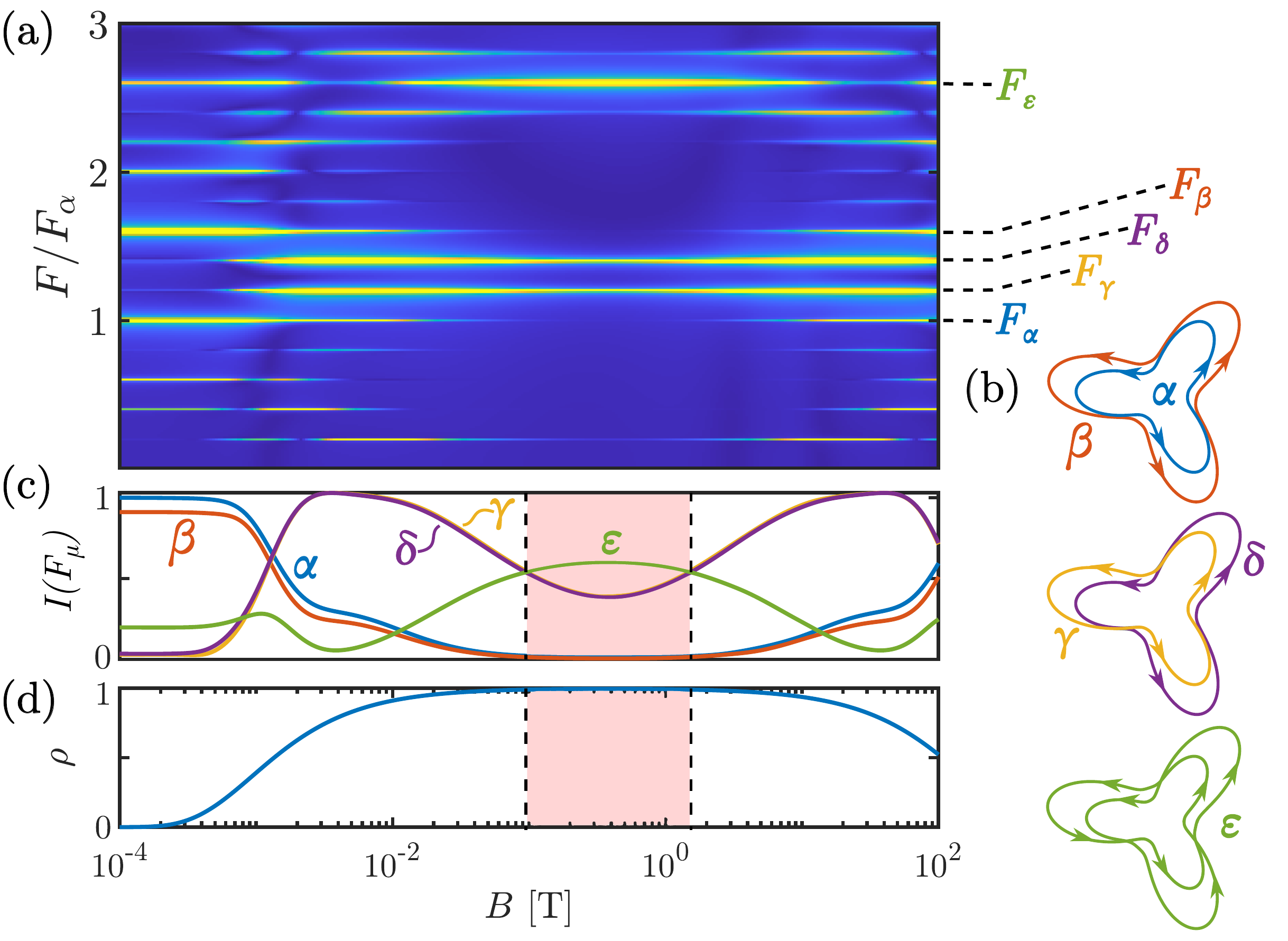}
    \caption{Magnetic field dependence of the quantum oscillation frequency spectrum (a), where we also indicate the frequencies $F_{\mu}$ corresponding to the five orbits $\mu=\alpha,\beta,\gamma,\delta$ depicted in (b). The intensities at these five frequencies and the tunnel probability are plotted in (c) and (d) as a function of magnetic field. In the red region, the M\"obius trajectory dominates. The effective parameters entering \equref{TransprobMainText} have been extracted from the continuum model for $\lambda_{\text{R}}=20\,\textrm{meV}$, $D_0=30\,\textrm{meV}$, and $\mu = 16.5\,\textrm{meV}$.}
    \label{fig:QuantumOscillations}
\end{figure}

Let us assume that $\mu$ is close to but not identical to $\mu^*$, leading to a finite but small minimal splitting $\delta=|\vec{g}_{\vec{k}_j^*}|$. In the adiabatic limit at small magnetic fields [$B \ll \delta^2/(e v_gv_F)$], the electrons simply follow the outer and inner Fermi surface, indicated as trajectory $\alpha$ and $\beta$ in \figref{fig:QuantumOscillations}(b). Meanwhile, finite $B$ will lead to a non-zero transition probability \cite{Supplement} 
\begin{equation}
    \rho=e^{-\pi \alpha_B}, \,\, \alpha_B = \frac{1}{2} \frac{\delta^2}{v_gv_F} \frac{1}{eB}\left[1 + \left(\frac{g\mu_B B}{2\delta}\right)^2 \right], \label{TransprobMainText}
\end{equation}
between the Fermi sheets at each of the three crossing points $\vec{k}=\vec{k}^*_j$ due to Landau-Zener tunneling \cite{SHEVCHENKO20101,PhysRevB.97.144422}. Here, $v_F$ is the Fermi velocity at $\vec{k}= \vec{k}^*_j$ and $v_g = |\partial_{k_\parallel} \vec{g}_{\vec{k}_j^*}|$, with $k_\parallel$ denoting the momentum along the Fermi surface. We can see in \figref{fig:QuantumOscillations}(d) that $\rho$ reaches a value close to $1$ already at moderately small magnetic fields, $B \gtrsim 10\,\textrm{mT}$, for the values of $v_g$, $v_F$, $\delta$ extracted from the continuum model at the indicated parameters. At very large magnetic fields, $B \gtrsim 10\,\textrm{T}$, the second term in \equref{TransprobMainText}, which describes the Zeeman-field-induced effective increase of the splitting between the inner and outer Fermi surfaces, starts to dominate and $\rho$ decreases again.
Using the frequently applied semi-classical approach \cite{RevModPhys.82.1959,PhysRevB.97.144422,PhysRevB.100.081405,PhysRevB.104.L201107,PhysRevLett.129.116804} and taking into account that intervalley scattering is typically negligible in clean moir\'e samples \cite{PabloQOs} due to the large momentum-space separation, we computed the Fourier spectrum of quantum oscillations shown in \figref{fig:QuantumOscillations}(a). As can be seen most clearly in \figref{fig:QuantumOscillations}(c), where the intensities at the frequencies $F_\mu$ associated with the fundamental trajectories are shown, there is a large magnetic field range, $0.1-2 \,\textrm{T}$, where the M\"obius trajectory $\epsilon$ dominates; more precisely, for this field range, which also significantly overlaps with the regime where quantum oscillations are most clearly visible in experiment \cite{PabloQOs}, the most prominent fundamental frequency corresponds to the \textit{sum} of the areas enclosed by the inner and outer Fermi surface---a hallmark signature of its M\"obius nature, requiring two revolutions to be closed. In fact, $\epsilon$ is visible and dominant over its constituent trajectories $\alpha$, $\beta$ in most of the experimentally accessible field range. Importantly, the experimental control over $\mu$ allows to adjust $\delta$ which, in turn, determines the magnetic field value $2\delta/g\mu_B$ where $\epsilon$ is most prominent; this should facilitate its successful experimental detection. 
We reiterate that for a M\"obius trajectory to dominate quantum oscillations, an odd number of crossing points is required. Setting aside quasi-crystals \cite{PhysRevB.100.081405} and the case without any rotation symmetry, the single-valley rotation symmetry $C^s_{3z}$ of the system is the only possibility to obtain an odd number of crossing points.

\vspace{1em}\textit{Superconductivity.}---We now examine the influence of SOC and $D_0$ on the superconducting state of tTLG, which we refer to as the {\it parent state}.  
The parent superconducting state comprises Cooper pairs formed from a partially-filled, spin-degenerate band. To be specific, we consider partial filling of the upper moir\'e bands, c.f. those of Fig. \ref{fig:Bandstructure}(a,b), where also superconductivity is observed experimentally \cite{Park2021_tTLG,HaoKim_tTLG,Jiang-Xiazi_diode,2021arXiv211207127S}. Moreover, we exclusively consider intervalley pairing since it is expected \cite{XuBalents2018, You2019, OurClassification,PhysRevB.106.104506} to be dominant over intravalley pairing due to time-reversal symmetry $\Theta_s$.  

To study the evolution of the superconducting state under combined SOC and $D_0$ we appeal to the linearized gap equation \cite{Supplement}, 
\begin{align}
\label{gapeq}
&d_{\mu, \bm k_1} = -\sum_{\bm k_2,\mu',\nu}\Gamma_{\mu\mu', \bm k_1,\bm k_2} {\cal W}_{\mu'\nu, \bm k_2} d_{\nu, \bm k_2},\\
\notag &{\cal W}_{\mu\nu,\bm k} = \sum_{n_{i},s_{j}}\frac{\tanh\left(\frac{\varepsilon_{+,n_1,\bm k}}{2T}\right)+\tanh\left(\frac{\varepsilon_{+,n_2,\bm k}}{2T}\right)}{2(\varepsilon_{+,n_1,\bm k}+\varepsilon_{+,n_2,\bm k})}\times\\
&\notag (s_\mu)_{s_2,s_3} C_{+,n_1, s_1, \bm k} C^*_{+,n_1, s_2, \bm k} C_{+,n_2, s_3, \bm k} C^*_{+,n_2, s_4, \bm k} (s_\nu)_{s_4,s_1}.
\end{align}
 The intervalley superconducting order parameter $d_{\mu,\vec{k}}$ encodes the momentum and spin structure, where $\mu=0$ and $\mu=1,2,3$ refer to the spin-singlet and triplet components, respectively. With perturbed Hamiltonian $h_\eta=h^{(g)}_\eta +h^{(t)}_\eta + h^{(\text{SOC})}_\eta+h^{(\text{D})}_\eta$,  
the factors  $C^*_{\eta, n, s, \bm k}\equiv\psi^\dagger_{\eta,n,\bm k} \psi^0_{\eta,s,\bm k}$ account for the projection of the perturbed eigenstates $h_\eta \psi_{\eta,n,\bm k}=\varepsilon_{\eta,n,\bm k}\psi_{\eta,n,\bm k}$ in band $n$ onto the unperturbed eigenstates $(h^{(g)}_\eta +h^{(t)}_\eta )\psi^0_{\eta,s,\bm k}=\varepsilon^0_{\eta,s,\bm k}\psi^0_{\eta,s,\bm k}$ of spin $s$, which comprise the parent superconducting state. We specialize to $\eta=+$ in \eqref{gapeq}; the $\eta=-$ order parameter follows from the fermion anticommutation relation.
Finally, the interaction vertex $\Gamma_{\mu,\nu, \bm k_1,\bm k_2}$ assumes an Anderson-Morel-type momentum structure, and encodes the spin-symmetry of the parent state \cite{Supplement}; we consider three cases: (i) SO(4), (ii) triplet-favored SO(3), and (iii) singlet-favored SO(3), with all states being invariant under $C_{3z}^s$ \cite{OurClassification}.

   \begin{figure}[t]
   \centering
    \includegraphics[width=1.0\linewidth]{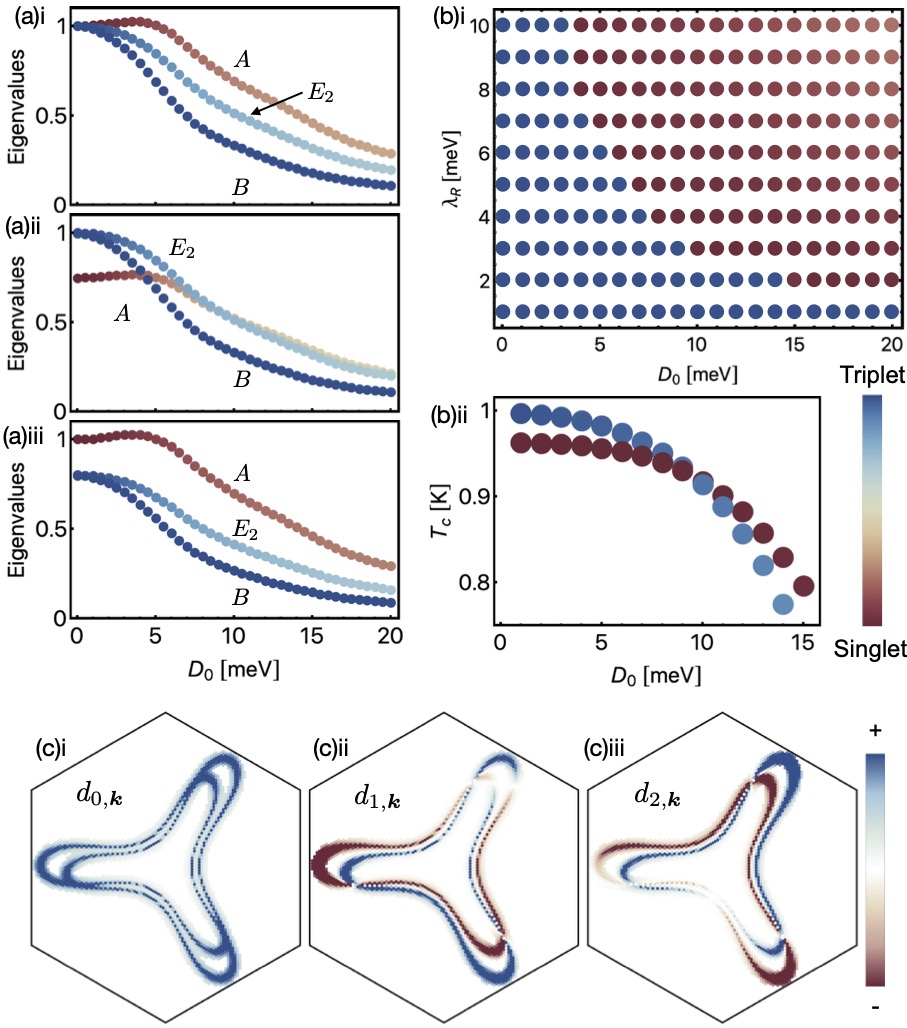}\vspace{0.2cm}
    \caption{Properties of the superconducting order. (a) Superconducting eigenvalues vs $D_0$ [normalized to unity at $D_0=0$] and at fixed $\lambda_{\text{R}}=30$ meV. The distinct IRs $(A, B, E_2)$ of $C_6^s$ are indicated. The three panels consider different spin-symmetries of the parent state: (ai) SO(4), (aii) triplet-favored SO(3) and (aiii) singlet-favored SO(3).  (bi) Phase diagram: colored points mark the dominant IR, with in-plane triplet $E_2$ (singlet $A$) state in blue (red). (bii) $T_c$ vs $D_0$ at $\lambda_{\text{R}}=5$ meV. (c) Order parameter of the leading $A$ state, decomposed into components $d_{\mu,\bm k}$; the $\mu=3$ component is zero due to $C_{2z}^s$ (not shown). Here $\theta=1.75^\circ$ and $\{\lambda_{\text{R}}, D_0\}=\{20,20\}$ meV [as per Fig.~\ref{fig:Bandstructure}(b,c,d)]. The $d_{0,\bm k}$ component transforms trivially under $C_{3z}^s$, while the admixed in-plane triplet $\bm d=(d_x, d_y)$ follows a M\"obius winding, with $\bm d_{\bm k} \parallel \bm g_{\bm k}$.}
    \label{fig:SC}
\end{figure} 

Having in mind $\theta_\text{TMD}\approx 30^\circ$, we focus on $\lambda_{\text{R}}\neq0$ and $\lambda_{\text{I}}=0$ here. When $D_0$ is turned on, the point symmetry is reduced to $C^s_6$, generated by six-fold out-of-plane rotations. It lacks $I$ and spin-rotation symmetry such that singlet and triplet can mix. More precisely, the parent triplet and singlet phases we consider separate into three distinct superconducting phases when $D_0\neq 0$, transforming under the irreducible representations (IRs) $A$, $B$, and $E_2$ of $C^s_6$. Here the $A$ state is predominantly spin-singlet, with admixed in-plane triplet; the $B$ state is pure out-of-plane triplet, since $C^s_{2z}$ prohibits any admixture of singlet or in-plane triplet components; and the $E_2$ state's two components are predominantly in-plane triplet, with admixed singlet. At fixed $\lambda_{\text{R}}\neq0$ we see in \figref{fig:SC}(a) that switching on the displacement field $D_0$ generates a splitting of these three distinct IRs. Hence, physically, the application of $D_0$ allows to change the nature of the superconducting state. 

Most strikingly, for the triplet favored parent state, $D_0$ drives a phase transition from $E_2$ pairing to an $A$ state, as signalled by the crossing of the respective eigenvalues in \figref{fig:SC}(a)(ii). This can be more clearly seen in \figref{fig:SC}(b)(ii), where we show the changes of the transition temperatures $T_c$ corresponding to the eigenvalues of these two states in the vicinity of the transition; both decrease with $D_0$ but $T_c$ decreases faster for the predominantly triplet state. As expected, the critical value of $D_0$ of this transition decreases with increasing $\lambda_{\text{R}}$, see \figref{fig:SC}(b)(i).

Finally, we turn to the spin and spatial structure of the superconducting state, encoded in $d_{\mu,\bm k}$. Figure \ref{fig:SC}(c) presents the leading state, $A$. Crucially, we see that the admixed in-plane triplet vector $\bm d_{\bm k}=(d_{1,\bm k},d_{2,\bm k})$ behaves as $\bm d_{\bm k}\parallel \bm g_{\bm k}$ and has a sign change between the Fermi surfaces; the superconducting order parameter $\bm d_{\bm k}$ thereby inherits the M\"obius winding around the Fermi surface.

\textit{Discussion and conclusion.}---We showed that the tunable SOC of tTLG with TMD layers on both sides allows to stabilize M\"obius-like Fermi surfaces, which result from vortices in the spin texture and are thus of a topological origin. We subsequently proposed quantum oscillations as a smoking gun probe. Turning to superconductivity, we showed that the parent superconducting state of tTLG is readily manipulated in our setup. As only a triplet parent state will give rise to a $D_0$-induced superconducting phase transition, this might provide a novel way to access the symmetry of the superconducting state of tTLG.
Our analysis can be readily extended to other correlated phases of tTLG, such as magnetic phases or nodal pairing states \cite{OurFollowUp} transforming non-trivially under $C_{3z}^s$ \cite{OurClassification}. In conjunction with experimental studies of the geometry in \figref{fig:overview}, we believe that this will allow probing the relevance of SOC systematically, e.g., for the stability of superconductivity \cite{TMDNadjPerge}, for the superconducting diode effect \cite{DiodeTheory,Jiang-Xiazi_diode}, or the recently observed microwave resonance \cite{2022arXiv220608354M} in twisted-graphene-TMD heterostructures.

\begin{acknowledgments}
H.D.S. thanks Zeb Krix and Sam Bladwell for useful discussions regarding quantum oscillations and acknowledges funding support from the Australian Research Council Centre of Excellence in Future Low-Energy Electronics Technology (FLEET) (CE170100039). M.S.S.~acknowledges funding from the European Union (ERC-2021-STG, Project 101040651---SuperCorr). Views and opinions expressed are however those of the authors only and do not necessarily reflect those of the European Union or the European Research Council. Neither the European Union nor the granting authority can be held responsible for them.

\end{acknowledgments}

\bibliography{draft_Refs}

\onecolumngrid
\newpage

\begin{appendix}

\section{Bandstructure and symmetries}\label{ContinuumModel}
In this appendix, we will first review the continuum model for tTLG, to be self-contained and establish notation. We will then add the proximity-induced SOC terms for the system geometry in \figref{fig:overview}, discuss the symmetries, and finally show more details on the resulting band structure. 

\hspace{1em}

\noindent\textbf{Continuum model description.} To capture the band structure of the system, we use a three-layer extension of the continuum model for twisted-bilayer graphene \cite{dos2007graphene,bistritzer2011moire,dos2012continuum,KhalafKruchkov2019,2019arXiv190712338L,PhysRevB.103.195411}. To write down the Hamiltonian, we define $c_{\vec{k};\rho,l,\eta,s,\vec{G}}$ to be the annihilation operator of an electron in graphene layer $l=1,2,3$ at momentum $\vec{k}$ in the moir\'e Brillouin zone (MBZ), in sublattice $\rho=A,B$, valley $\eta = \pm$, with spin $s=\uparrow,\downarrow$ and reciprocal moir\'e lattice (RML) vector $\vec{G} = \sum_{j} n_j \vec{G}_j$, $n_j \in \mathbbm{Z}$. Throughout this work, we use the same symbol for Pauli matrices and the associated quantum numbers; as such, $\rho_j$ are Pauli matrices in sublattice, $s_j$ in spin, and $\eta_j$ in valley space, where $j=0,1,2,3$.
To make the mirror symmetry $\sigma_h$ of the system more apparent, we go to its eigenbasis \cite{KhalafKruchkov2019} by introducing the field operators $\psi_{\vec{k};\rho,\ell,\eta,s,\vec{G}}$ as
\begin{equation}
    c_{\vec{k};\rho,l,\eta,s,\vec{G}} = V_{l,\ell} \psi_{\vec{k};\rho,\ell,\eta,s,\vec{G}}, \qquad V=\frac{1}{\sqrt{2}} \begin{pmatrix} 1 & 0 & -1 \\ 0 & \sqrt{2} & 0 \\ 1 & 0 & 1 \end{pmatrix}, \label{TrafoToMirrorEigenbasis}
\end{equation}
where $\ell = 1,2$ and $\ell = 3$ refer to the mirror-even and mirror-odd subspaces, respectively. 

We split the continuum Hamilton,
\begin{equation}
    H_0^{\text{Full}} = \sum_{\vec{k} \in \text{MBZ}} \sum_{\rho,\rho'=A,B} \sum_{\ell,\ell'=1,2,3} \sum_{\eta=\pm} \sum_{s,s'=\uparrow,\downarrow}  \sum_{\vec{G},\vec{G}' \in \text{RL}} \psi^\dagger_{\vec{k};\rho,\ell,\eta,s,\vec{G}} \left(h_{\vec{k},\eta}\right)_{\rho,\ell,s,\vec{G};\rho',\ell',s',\vec{G}'} \psi^\pdagger_{\vec{k};\rho',\ell',\eta,s',\vec{G}'}, \label{FullContinuumModel}
\end{equation}
into four parts, $h_{\vec{k},\eta} = h^{(g)}_{\vec{k},\eta} + h^{(t)}_{\vec{k},\eta} + h^{(D)}_{\vec{k}} + h^{(\text{SOC})}_{\vec{k},\eta}$, corresponding to the contribution of the individual graphene layers, the tunneling between the layers, the coupling of the electric displacement field, and the proximity-induced spin-orbit coupling terms.
The first term reads as
\begin{align}
    \left(h^{(g)}_{\vec{k},+}\right)_{\rho,\ell,s,\vec{G};\rho',\ell',s',\vec{G}'} &= \delta_{\ell,\ell'} \delta_{s,s'}\delta_{\vec{G},\vec{G}'} v_F (\vec{\rho}_{\theta_\ell})_{\rho,\rho'} \left(\vec{k} + \vec{G} - (-1)^\ell \vec{q}_{1}/2 \right), \label{DiracCones} \\ \left(h^{(g)}_{\vec{k},-}\right)_{\rho,\ell,s,\vec{G};\rho',\ell',s',\vec{G}'} &= \left(h^{(g)}_{-\vec{k},+}\right)^*_{\rho,\ell,s,-\vec{G};\rho',\ell',s',-\vec{G}'},
\end{align}
where $\vec{\rho}_{\theta} = e^{i \theta \rho_z/2} \vec{\rho} e^{-i \theta \rho_z/2}$, and $\vec{q}_1$ connects the K and K' points of the MBZ. For the tunneling between the layers, we take the common form
\begin{align}\begin{split}
    \left(h^{(t)}_{\vec{k},+}\right)_{\rho,\ell,s,\vec{G};\rho',\ell',s',\vec{G}'} &= \sqrt{2}\, \delta_{s,s'} \hspace{-0.2em} \begin{pmatrix} 0 & (T_{\vec{G}-\vec{G}'})_{\rho,\rho'} & 0 \\  (T_{\vec{G}'-\vec{G}}^*)_{\rho',\rho} & 0 & 0 \\ 0 & 0 & 0 \end{pmatrix}_{\ell,\ell'}, \hspace{-0.3em} \\ \left(h^{(t)}_{\vec{k},-}\right)_{\rho,\ell,s,\vec{G};\rho',\ell',s',\vec{G}'} &= \left(h^{(t)}_{-\vec{k},+}\right)^*_{\rho,\ell,s,-\vec{G};\rho',\ell',s',-\vec{G}'}. \label{TunnelingMatrixNotation}
\end{split}\end{align}
with
\begin{align}
    T_{\delta\vec{G}} = \sum_{j=-1,0,1}\delta_{\delta\vec{G}+\vec{A}_j,0} \left[w_0 \rho_0 + w_1 \begin{pmatrix} 0 & \omega^j \label{FormOfT} \\  \omega^{-j} & 0 \end{pmatrix} \right], \\ \omega = e^{i \frac{2\pi}{3}}, \quad \vec{A}_0 =0, \quad \vec{A}_1 = \vec{G}_1, \quad \vec{A}_2 = \vec{G}_1 + \vec{G}_2, 
\end{align}
which only keeps the lowest-order reciprocal lattice vectors in the Fourier expansion. A finite displacement field leads to the contribution
\begin{equation}
    \left(h^{(D)}_{\vec{k}}\right)_{\rho,\ell,s,\vec{G};\rho',\ell',s',\vec{G}'} = -D_0 \delta_{\rho,\rho'}\delta_{s,s'}\delta_{\vec{G},\vec{G}'} \begin{pmatrix} 0 & 0 & 1 \\ 0 & 0 & 0 \\ 1 & 0 & 0 \end{pmatrix}_{\ell,\ell'}. \label{MatrixFormOfD}
\end{equation}
Note that the inter-layer tunneling preserves the mirror symmetry, which is why it does not act between the $\ell=1,2$ (even) and $\ell=3$ (odd) sectors in \equref{TunnelingMatrixNotation}, while $D_0$ is odd under it and therefore is entirely off-diagonal in the two sectors, as can be seen in \equref{MatrixFormOfD}.

\begin{table*}[tb]
\begin{center}
\caption{Action of the point symmetries of the continuum theory on the microscopic field operators $\psi_{\vec{k}}$ defined in \equref{TrafoToMirrorEigenbasis} and on the low-energy operators $c_{\vec{k}}$ describing the band structure of the bands that cross the Fermi level via the effective Hamiltonian $H^{\text{eff}} = \sum_{\vec{k}} c^\dagger_{\vec{k},\eta} (s_0 \xi_{\eta\cdot\vec{k}} + \eta\, \vec{g}_{\eta\cdot\vec{k}}\cdot\vec{s}) c^\pdagger_{\vec{k},\eta}$. For convenience of the reader, we also list redundant symmetries. As we are primarily interested in $\lambda_{\text{R}}, \lambda_{\text{I}} \gg \lambda_{\text{KM}}, m$ \cite{PhysRevB.104.195156,PhysRevB.99.075438,PhysRevB.100.085412,2022arXiv220609478L,PhysRevResearch.4.L022049} and for simplicity, we only state the constraints on $D_0, \lambda_{\text{R}}$, and $\lambda_{\text{I}}$ for the respective symmetry to be present in the last column while assuming $\lambda_{\text{KM}}, m = 0$.}
\label{ActionOfSymmetries}
\begin{ruledtabular}
 \begin{tabular} {ccccc} 
Symmetry $S$ & unitary?  & $S\psi_{\vec{k};\ell,\vec{G}}S^\dagger$ & $S c_{\vec{k}}S^\dagger$ & condition \\ \hline
SO(3)$_{s}$ & \cmark    & $e^{i\vec{\varphi}\cdot \vec{s}}  \psi_{\vec{k};\ell,\vec{G}}$  & $e^{i\vec{\varphi}\cdot \vec{s}}  c_{\vec{k}}$ &  $\lambda_{\text{R}}=\lambda_{\text{I}}=0$  \\ 
SO(2)$_{s}$ & \cmark    & $e^{i \varphi s_z}  \psi_{\vec{k};\ell,\vec{G}}$ & $e^{i\varphi s_z}  c_{\vec{k}}$  & $\lambda_{\text{R}}=0$  \\ \hline
$C_{3z}$ & \cmark    & $e^{i\frac{2\pi}{3} \rho_z\eta_z}  \psi_{C_{3z}\vec{k};\ell,C_{3z}\vec{G}}$  & $c_{C_{3z}\vec{k}}$ & $\lambda_{\text{R}}=0$  \\
$C^s_{3z}$ & \cmark    & $e^{i\frac{2\pi}{3} (\rho_z\eta_z + s_z)}  \psi_{C_{3z}\vec{k};\ell,C_{3z}\vec{G}}$ & $e^{i\frac{2\pi}{3} s_z}c_{C_{3z}\vec{k}}$ & ---  \\ \hline
$C_{2z}$ & \cmark   & $\eta_x \rho_x \psi_{-\vec{k};\ell,-\vec{G}}$ & $\eta_x c_{-\vec{k}}$ & $\lambda_{\text{R}}=\lambda_{\text{I}}=0$   \\
$C^s_{2z}$ & \cmark   & $s_z\eta_x \rho_x \psi_{-\vec{k};\ell,-\vec{G}}$ & $s_z\eta_x c_{-\vec{k}}$ & $\lambda_{\text{I}}=0$  \\
$C^{s'}_{2z}=C^s_{2z} i  s_{y,x}$ & \cmark   & $s_{x,y}\eta_x \rho_x \psi_{-\vec{k};\ell,-\vec{G}}$ & $s_{x,y}\eta_x c_{-\vec{k}}$ & $\lambda_{\text{R}}=0$  \\ \hline
$\sigma_h$ & \cmark   & $ (1,1,-1)_\ell \psi_{\vec{k};\ell,\vec{G}}$ & $\pm c_{\vec{k}}$  & $D_0=\lambda_{\text{R}}=\lambda_{\text{I}}=0$  \\
$\sigma^s_h$ & \cmark    & $s_z(1,1,-1)_\ell \psi_{\vec{k};\ell,\vec{G}}$ & $\pm s_z c_{\vec{k}}$  & $D_0=\lambda_{\text{I}}=0$  \\
$\sigma^{s'}_h = \sigma^s_h i s_{y,x}$ & \cmark   & $s_{x,y}(1,1,-1)_\ell \psi_{\vec{k};\ell,\vec{G}}$ &  $\pm s_{x,y} c_{\vec{k}}$ & $D_0=\lambda_{\text{R}}=0$  \\ \hline
$I = C_{2z}\sigma_h = C^s_{2z}\sigma^s_h$ & \cmark    & $ \eta_x\rho_x(1,1,-1)_\ell \psi_{-\vec{k};\ell,\vec{G}}$ & $\pm \eta_x c_{-\vec{k}}$ & $D_0=0$  \\ \hline
$\Theta$ & \xmark    & $\eta_x \psi_{-\vec{k};\ell,-\vec{G}}$ & $\eta_x c_{-\vec{k}}$ & $\lambda_{\text{R}}=\lambda_{\text{I}}=0$  \\
$\Theta^s$ & \xmark   & $is_y\eta_x \psi_{-\vec{k};\ell,-\vec{G}}$ & $is_y\eta_x c_{-\vec{k}}$  & ---  \\
 \end{tabular}
\end{ruledtabular}
\end{center}
\end{table*}

Finally, let us take into account the additional terms induced by the proximity of the outer two layers to a TMD material, see \figref{fig:overview}. To construct the corresponding term, $h^{(\text{SOC})}_{\vec{k}}$, in the Hamiltonian, we start from its form in single-layer graphene \cite{Gmitra2015},
\begin{align}
    H_{0,l=1}^{\text{SOC}} &= \sum_{\vec{k} \in \text{MBZ}} \sum_{\rho,\rho'=A,B}  \sum_{\eta=\pm} \sum_{s,s'=\uparrow,\downarrow}  \sum_{\vec{G}\in \text{RL}} c^\dagger_{\vec{k};\rho,1,\eta,s,\vec{G}} \left(h^{\text{SOC},l=1}_{\eta}\right)_{\rho,s;\rho',s'} c^\pdagger_{\vec{k};\rho',1,\eta,s',\vec{G}}, \\
    h^{\text{SOC},l=1}_{\eta} &=\lambda_{\text{I}} s_z \eta + \lambda_{\text{R}} e^{-i \phi s_z/2} \left(\eta \rho_x s_y - \rho_y s_x \right) e^{i \phi s_z/2} + \lambda_{\text{KM}} \eta \rho_z s_z + m \rho_z. \label{SOCLayer1}
\end{align}
The first term, proportional to $\lambda_{\text{I}}$, describes ``Ising'' SOC (since it is still invariant under a residual SO(2) spin rotation), the second ($\lambda_{\text{R}}$) is the ``Rashba'' term $\lambda_{\text{R}}$, and the third one ($\lambda_{\text{KM}}$) is usually referred to as ``Kane-Mele'' \cite{PhysRevLett.95.226801} SOC. Among all terms in \equref{SOCLayer1}, the Kane-Mele term is the only one symmetric under two-fold spinless out-of-plane rotation $C_{2z}$ (with representation $\eta_x\rho_x$, $\vec{k}\rightarrow -\vec{k}$ in single-layer graphene), while all other three are odd under it.
The spin-angle $\phi$, which depends \cite{PhysRevB.104.195156,2022arXiv220609478L,PhysRevResearch.4.L022049} on the twist angle $\theta_{\text{TMD}}$ between graphene and the TMD layer, can be set to $\phi=0$ without loss of generality for our analysis, since we can absorb it by performing a global spin rotation along the $s_z$ axis.  Finally, $m$ captures the fact that the TMD layer breaks $C_{2z}$ such that an imbalance of graphene's sublattices is allowed by symmetry. 

All parameters $\lambda_{\text{I}}$, $\lambda_{\text{R}}$, $\lambda_{\text{KM}}$, and $m$ have been computed in a series of first-principle works \cite{PhysRevB.104.195156,PhysRevB.99.075438,PhysRevB.100.085412,2022arXiv220609478L,PhysRevResearch.4.L022049} and are found to depend crucially on $\theta_{\text{TMD}}$ and the TMD material. Importantly, the TMD-graphene bilayer system exhibits a mirror symmetry at $\theta_{\text{TMD}}=\pi/6$, which implies that both $m$ and $\lambda_{\text{I}}$ vanish. While it is straightforward to also include finite $m$ and $\lambda_{\text{KM}}$ in our analysis, we focus on $m=\lambda_{\text{KM}}=0$ in the main text to keep the discussion concise which is justified as follows: $\lambda_{\text{KM}}$ is universally found to be negligibly small for any $\theta_{\text{TMD}}$ and also $m$ is typically much smaller than $\lambda_{\text{I}}$ and $\lambda_{\text{R}}$, in particular for  $\pi/12<\theta_{\text{TMD}} <\pi/6$ \cite{PhysRevB.104.195156,PhysRevB.99.075438,PhysRevB.100.085412,2022arXiv220609478L}; furthermore, \refcite{PhysRevB.99.075438} argued that $m$ and $\lambda_{\text{KM}}$ are even required to vanish in an incommensurate TMD-graphene superlattice.

\begin{figure}[tb]
   \centering
    \includegraphics[width=0.7\linewidth]{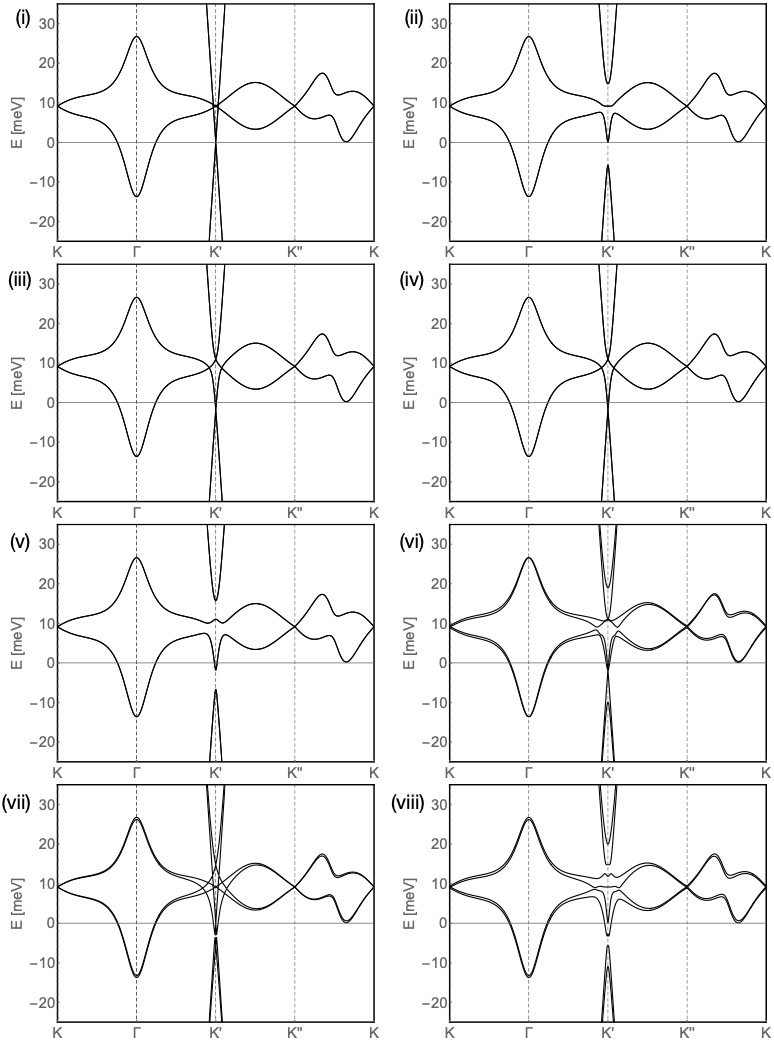}
    \caption{Band structure at twist $\theta=1.75^\circ$ for the following combinations of SOC and $D_0$: (i) $\lambda_{\text{R}}, \lambda_{\text{I}}, D_0 = \{0,0,0\}$meV,  (ii) $\lambda_{\text{R}}, \lambda_{\text{I}}, D_0 = \{10,0,0\}$meV, (iii) $\lambda_{\text{R}}, \lambda_{\text{I}}, D_0 = \{0,10,0\}$meV, (iv) $\lambda_{\text{R}}, \lambda_{\text{I}}, D_0 = \{0,0,10\}$meV, (v) $\lambda_{\text{R}}, \lambda_{\text{I}}, D_0 = \{10,10,0\}$meV, (vi) $\lambda_{\text{R}}, \lambda_{\text{I}}, D_0 = \{10,0,10\}$meV, (vii) $\lambda_{\text{R}}, \lambda_{\text{I}}, D_0 = \{0,10,10\}$meV, (viii) $\lambda_{\text{R}}, \lambda_{\text{I}}, D_0 = \{10,10,10\}$meV}
    \label{fig:BS}
\end{figure}

With $h^{\text{SOC},l=1}_{\eta}$ in \equref{SOCLayer1} being the coupling in layer $l=1$, the coupling in layer $l=3$ follows by inversion symmetry, $I=C_{2z}\sigma_h$,
\begin{align}
    H_{0,l=3}^{\text{SOC}} &= \sum_{\vec{k} \in \text{MBZ}} \sum_{\rho,\rho'=A,B}  \sum_{\eta=\pm} \sum_{s,s'=\uparrow,\downarrow}  \sum_{\vec{G}\in \text{RL}} c^\dagger_{\vec{k};\rho,3,\eta,s,\vec{G}} \left(h^{\text{SOC},l=3}_{\eta}\right)_{\rho,s;\rho',s'} c^\pdagger_{\vec{k};\rho',3,\eta,s',\vec{G}}, \\
    h^{\text{SOC},l=3}_{\eta} &= \rho_x h^{\text{SOC},l=1}_{-\eta} \rho_x = -\lambda_{\text{I}} s_z \eta - \lambda_{\text{R}}  \left(\eta \rho_x s_y - \rho_y s_x \right)  + \lambda_{\text{KM}} \eta \rho_z s_z - m \rho_z.
\end{align}
We assume that there is no direct proximity coupling to the middle layer and, hence, $H_{0,l=2}^{\text{SOC}}$=0. Upon transforming to the mirror eigenbasis according to \equref{TrafoToMirrorEigenbasis}, we find
\begin{equation}
    \left(h^{(\text{SOC})}_{\vec{k},\eta}\right)_{\rho,\ell,s,\vec{G};\rho',\ell',s',\vec{G}'} = \delta_{\vec{G},\vec{G}'} \begin{pmatrix} (h_\eta^{\text{s}})_{\rho,s;\rho',s'} & 0 & (h_\eta^{\text{a}})_{\rho,s;\rho',s'} \\ 0 & 0 & 0 \\ (h_\eta^{\text{a}})_{\rho,s;\rho',s'} & 0 & (h_\eta^{\text{s}})_{\rho,s;\rho',s'} \end{pmatrix},
\end{equation}
where we define the mirror symmetric (s) and anti-symmetric (a) combinations of the individual layer SOC coupling terms,
\begin{align}
    h_\eta^{\text{s}} &= \frac{1}{2}\left( h^{\text{SOC},l=1}_{\eta} + h^{\text{SOC},l=3}_{\eta}\right) = \lambda_{\text{KM}} \eta \rho_z s_z \\
    h_\eta^{\text{a}} &= \frac{1}{2}\left( h^{\text{SOC},l=3}_{\eta} - h^{\text{SOC},l=1}_{\eta} \right) = -\lambda_{\text{I}} s_z \eta - \lambda_{\text{R}} \left(\eta \rho_x s_y - \rho_y s_x \right) - m \rho_z
\end{align}
As required by symmetry, the mirror-symmetric (anti-symmetric) combinations appear only on the diagonals within each (off-diagonals between different) mirror-symmetry sectors. This is in line with the observation that the spin-orbit splitting of the Fermi surfaces in \figref{fig:Bandstructure}(d) tends to be larger in the regions closer to the $K'$ point since, for the valley shown, the mirror-even and mirror-odd sectors are closer in energy around $K'$ [see \figref{fig:Bandstructure}(b)].
With this, we have defined all parts of the continuum-model Hamiltonian $H_0^{\text{Full}}$ in \equref{FullContinuumModel}.

\hspace{1em}

\noindent\textbf{Single-particle symmetries.} As it is important for our discussion of superconducting instabilities, we list the point symmetries of the continuum-model Hamiltonian (\ref{FullContinuumModel}) and their representations in \tableref{ActionOfSymmetries}. While there are additional approximate internal symmetries, these are not crucial for our analysis and, hence, not included (we refer to \refcite{OurPRXTTLG} where these are discussed for tTLG without SOC, using a notation similar to our notation here). To keep the notation compact, we use the additional superscript $s$ to indicate that the operator involves a spin-space transformation on top of its action in all other quantum numbers---momentum $\vec{k}$, sublattice, layer, and valley. For instance, $C_{3z}$ is just a three-fold rotation along the $z$ direction, leaving the spin unchanged, while $C^s_{3z}$ also rotates the spin as indicated. 

Most importantly for our analysis: at $D_0=0$ there is a $\vec{k}$-local anti-unitary symmetry,
\begin{equation}
    I\Theta_s=C_{2z}\sigma_h \Theta_s = \rho_x \begin{pmatrix}1 & 0 & 0 \\ 0 & 1 & 0 \\ 0 & 0 & -1\end{pmatrix} is_y \mathcal{K}, \label{AntiUnitarySymmetryDoubleDeg}
\end{equation}
where the $3\times 3$ matrix acts in pseudo-layer (or mirror-eigenvalue) space, with label $\ell$ in \equref{TrafoToMirrorEigenbasis}, and $\mathcal{K}$ denotes complex conjugation.
Importantly, $C_{2z}\sigma_h \Theta_s$ squares to $-1$ and, hence, guarantees the presence of pseudo-spin-degenerate bands. Obviously, this symmetry is broken at $D_0\neq 0$, thus, removing the pseudo-spin degeneracy. As such, we can use $D_0$ to turn on and off as well as tune the strength of the spin-orbit splitting of the bands, which can be explicitly seen in \figref{fig:Bandstructure} of the main text and \figref{fig:BS} below.

\hspace{1em}

\noindent\textbf{Band structure plots.} Figure \ref{fig:BS} presents the band structure at twist angle $\theta=1.75^\circ$, for various combinations of the SOC parameters $\lambda_{\text{R}}, \lambda_{\text{I}}$ and displacement field value $D_0$. This illustrates the following expectations based on the symmetries discussed above: first, only if \textit{both} $D_0$ and SOC are finite will the spin degeneracy of the bands be removed---SOC alone is not enough due to the anti-unitary symmetry (\ref{AntiUnitarySymmetryDoubleDeg}). Second, $\sigma_h$ will be broken if one or more of $\{\lambda_{\text{R}}, \lambda_{\text{I}}, D_0\}$ is non-zero, leading to an admixture of the twisted-bilayer-graphene (mirror even) and graphene (mirror-odd) sectors of tTLG graphene.

\section{Quantum oscillations}\label{AppendixQO}
In this section, we first provide more details on how the results of \figref{fig:QuantumOscillations} of the main text are obtained and then explicitly contrast it with the quantum oscillations when the number of avoided crossing points is even.

\begin{figure}[t]
   \centering
    \includegraphics[width=0.9\linewidth]{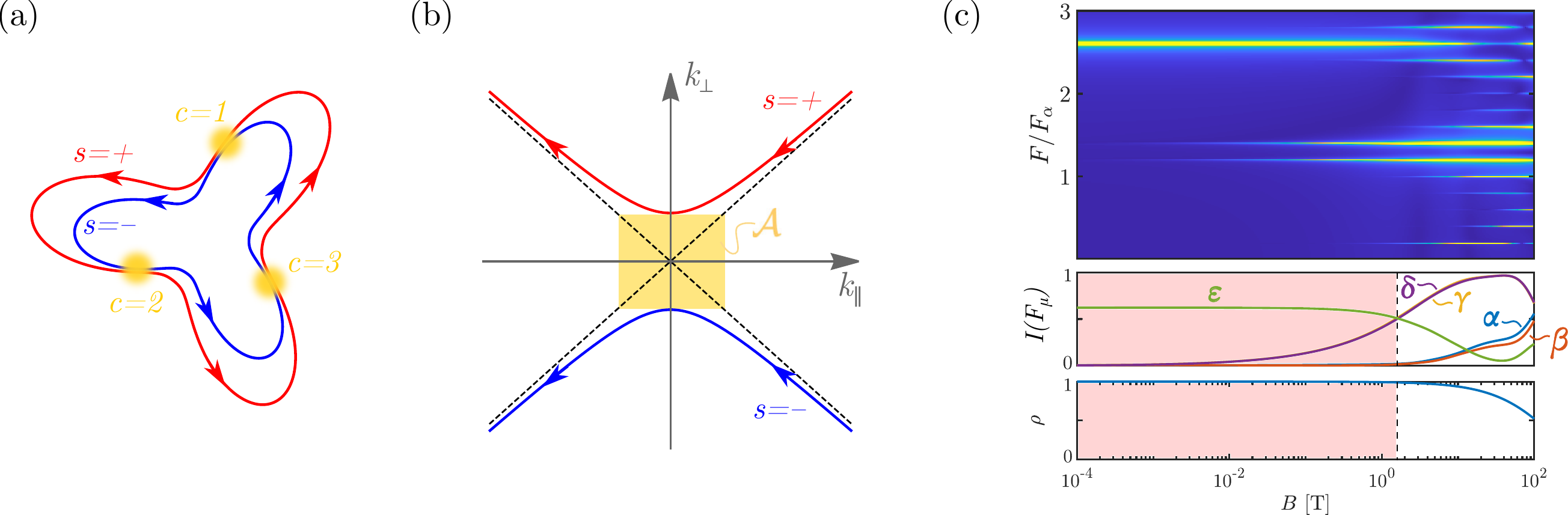}
    \caption{For our semi-classical approach, the Fermi surfaces (red and blue lines) are split into $N_c$ segments separated by scattering regions $c=1,2,\dots N_c$ (yellow) as shown in (a) for the for us relevant case of $N_c=3$. In (b), we display the Fermi surfaces in the immediate vicinity of a scattering region as described by the effective model in \equref{EffectiveModel}. The area $\mathcal{A}$ of the yellow rectangle determines the transition probability via \equref{ExpressionForAlphaB}. (c) is the same as \figref{fig:QuantumOscillations}(a,c,d) with $\mu$ fine-tuned to $\mu^*$ such that the M\"obius trajectory also dominates for small magnetic fields.}
    \label{fig:transitionprob}
\end{figure}

We here employ a commonly used \cite{RevModPhys.82.1959,PhysRevB.97.144422,PhysRevB.100.081405,PhysRevB.104.L201107,PhysRevLett.129.116804} semi-classical theory that allows to capture the interference effects between the electron trajectories on the outer (red, label $s=+$) and inner (blue, label $s=-$) Fermi surfaces in \figref{fig:transitionprob}(a). Let us denote the ($2\times 2$) scattering matrix associated with the avoided crossings $c=1,2,\dots N_c$ by $\mathcal{S}^c_{s,s'}$ and let the phases accumulated on the parts of the Fermi surfaces connecting the scattering regions $c$ and $c'$ be on the diagonal of the diagonal matrix $\Lambda^{c\rightarrow c'} = \text{diag}(e^{i\Omega_+^{c\rightarrow c'}},e^{i\Omega_-^{c\rightarrow c'}})$. Then, demanding that the time evolution of the amplitudes of an electron at the Fermi surfaces be single-valued implies
\begin{equation}
    \det \left(U-\mathbbm{1}\right) = 0\quad \text{with} \quad U = \prod_{c=1}^{N_c} \mathcal{S}^c \Lambda^{c\rightarrow c+1 \text{mod} N_c}, \label{QuantizationCondition}
\end{equation}
which provides a quantization condition. Throughout this work, we will assume that all segments $c\rightarrow c+1$ and avoided crossings are related by a $C_{N_c}$ symmetry and choose a gauge where $\mathcal{S}^c = \mathcal{S}$ and $\Lambda^{c\rightarrow c+1 \text{mod} N_c} = \Lambda$, independent of $c$; in that case, we simply have $U=(\mathcal{S}\Lambda)^{N_c}$.

\vspace{1em}\noindent\textbf{Scattering matrix.} The scattering matrix for interband magnetic breakdown is given by \cite{PhysRevB.97.144422} 
\begin{equation}
    S = \begin{pmatrix} \sqrt{1-\rho^2}e^{i\omega(\alpha_B)} & -\rho e^{i\vartheta} \\ \rho e^{-i\vartheta}& \sqrt{1-\rho^2}e^{-i\omega(\alpha_B)} \end{pmatrix}, \label{GlazmanScattering}
\end{equation}
where $\rho$ is the Landau-Zener-like tunneling probability, 
\begin{equation}
    \rho(B)=e^{-\pi \alpha_B}, \quad \alpha_B=\frac{1}{8}\mathcal{A}l_{B}^2, \quad l_{B} = \frac{1}{\sqrt{eB}}, \label{ExpressionForAlphaB}
\end{equation}
with $\mathcal{A}$ denoting the area of the rectangle enclosed by the two branches of the Fermi surfaces, see \figref{fig:transitionprob}(b), and $B$ the magnetic field. Furthermore, we have $\omega(\alpha_B) = \alpha_B(1-\ln \alpha_B) + \text{arg}\Gamma(i\alpha_B) + \pi/4$ in \equref{GlazmanScattering}, where $\Gamma$ is the Gamma function. Finally, $\vartheta$ is the phase related to the relative U(1)-gauge freedom between wavefunctions in the two spin-orbit split bands; it is straightforward to check that it will drop out \cite{PhysRevB.97.144422} for closed loops and therefore in the quantization condition (\ref{QuantizationCondition}), as required by gauge invariance. 

Importantly, since the states on the two Fermi surfaces that we study have different spin-polarizations, a magnetic field affects $\alpha_B$ not only via the (orbital) magnetic length $l_B$ but also via the Zeeman field which changes $\mathcal{A}$ in \equref{ExpressionForAlphaB}. To study this quantitatively, let us consider a single avoided crossing and denote the deviation of the momentum from the avoided crossing parallel (perpendicular) to the Fermi surfaces right at the avoided crossing point by $k_\parallel$ ($k_\perp$), see \figref{fig:transitionprob}(b). We can then write the effective Hamiltonian in the vicinity of the crossing as \footnote{We have checked that adding momentum dependence according to $\delta \rightarrow \delta + v_g k_\perp$ in \equref{EffectiveModel} does not change our results for the parameters used in this work. We thus neglect it here for simplicity of the presentation.}
\begin{equation}
    h^{\text{eff}}(\vec{k}) \sim \sigma_0 \,v_F k_\perp + (\sigma_x,\sigma_y)^T R(\varphi) (v_g k_\parallel,\delta ) - \frac{g\mu_B}{2}  B \sigma_z, \label{EffectiveModel}
\end{equation}
where $v_F$ is the Fermi velocity at the crossing point, $R(\varphi)$ is an O(2) rotation matrix determining the local orientation of the (pseudo)-spin quantum number, $v_g$ is the velocity describing the change of the spin-orbit vector with $k$, and $\delta$ its magnitude right at the crossing point ($k_\parallel=0$), i.e., the minimal energy separation of the spin-orbit-split bands for $B=0$. Finally, $\mu_B = \frac{e}{2m_e}$ is the Bohr magneton and $g= 2$ the electron's Land\'e factor.
Computing $\mathcal{A}$ for this model, we get from \equref{ExpressionForAlphaB}
\begin{equation}
    \rho(B)=e^{-\pi \alpha_B}, \quad \alpha_B = \frac{1}{2} \frac{\delta^2}{v_gv_F} \frac{1}{eB}\left[1 + \left(\frac{g\mu_B B}{2\delta}\right)^2 \right]. \label{TransitionProb1}
\end{equation}
As can be seen in \figref{fig:QuantumOscillations}(d), this leads to a non-monotonic behavior of $\alpha_B$ and, thus, $\rho$ with magnetic field: upon increasing $B$, $\alpha_B$ first decreases and reaches its minimum at
\begin{equation}
    B_0 = \frac{2\delta}{g\mu_B } \qquad \text{where} \quad \alpha_B|_{B=B_0} = \frac{\delta}{2m_e v_gv_F} =: \alpha_0 ,
\end{equation}
before increasing again due to the Zeeman-induced increase of the gap. With this, we can write \equref{TransitionProb1} in the compact dimensionless form
\begin{equation}
    \rho(B) = e^{-\frac{\pi}{2} \alpha_0  (b+b^{-1})  }, \qquad b = B/B_0.
\end{equation}
By fitting the effective Hamiltonian in \equref{EffectiveModel} to our bandstructure calculations, we find $\delta \approx 0.022\,\textrm{meV}$, $v_F\approx 0.076 v_0$, and $v_g\approx 0.02 v_0$, with $v_0$ being the velocity of single-layer graphene, for $\lambda_{\text{R}}=20\,\textrm{meV}$, $D=30\,\textrm{meV}$, and $\mu = 16.5\,\textrm{meV}$; this leads to $B_0 \approx 0.38\,\textrm{T}$ and $\alpha_0 = 0.0016$, i.e., a maximum tunnel probability of about $0.995$. These are the parameters used in \figref{fig:QuantumOscillations}(a,c,d) of the main text.

We note, however, that the value of $\delta$ can be tuned in experiment by gate voltage as it crucially depends on the difference of the chemical potential $\mu$ and the energy $\mu^*$ of the vortices in \figref{fig:Bandstructure}(c). Specifically, by tuning $\mu \rightarrow \mu^*$, we obtain $\delta \rightarrow 0$ and, thus,
\begin{equation}
    \rho(B) \sim e^{-\pi B/B^*}, \quad B^* = 2\left(\frac{2}{g}\right)^2 \frac{v_g v_F e}{\mu_B^2}.
\end{equation}
For the parameters above we find $B^* \approx 480\,\textrm{T}$. The associated transition probability looks virtually identical to that shown in \figref{fig:QuantumOscillations}(d) of the main text to the right of the red region and indistinguishable from $1$ to the left of it. Consequently, this pushes the lower bound of the magnetic-field range where the M\"obius trajectory dominates to zero, as can be seen in \figref{fig:transitionprob}(c).

\vspace{1em}\noindent\textbf{Phases on Fermi segments.} For the phase conventions introduced above, the phases accumulated on each of the $C_{N_c}$-related outer ($s=+$) and inner ($s=-$) Fermi-surface segments between consecutive scattering regions are given by \cite{PhysRevB.97.144422}
\begin{equation}
    \Omega_s = \frac{A_s l_B^2}{N_c} + \frac{\pi}{N_c} + \phi^s_{\text{Berry}}(B), \qquad s = \pm, \label{PhasesAccumulated}
\end{equation}
where $A_s$ is the oriented momentum-space area enclosed by the outer $s=+$ (inner $s=-$) Fermi surface. The second term on the right-hand side results from the Maslov phase \cite{KELLER1958180} for our Fermi surfaces that are deformable into a circle, and the last is the Berry phase contribution for that segment, to be discussed shortly. As a simple consistently check, we first note that the resonance condition (\ref{QuantizationCondition}) becomes equivalent to
\begin{equation}
    A_s l_B^2 = (2l+1)\pi, \qquad l \in \mathbbm{Z},
\end{equation}
for $s=+$ or $s=-$, in the limit without any interband tunneling ($\rho=0$) and setting $\phi^s_{\text{Berry}}=0$. This is the classic result of Onsager, Lifshitz, and Kosevich \cite{Onsager,lifshitz1956theory}. 

For a spin texture like the one shown in \figref{fig:Bandstructure}(d), we have $\phi^s_{\text{Berry}}(B=0)=2\pi/N_c$ (here with $N_c=3$). A finite magnetic field, cants the spin out of plane such that $N_c\phi^s_{\text{Berry}}(B=0)$ is not an integer multiple of $\pi$. However, the characteristic magnetic field scale where the canting and, thus, the impact on the Berry phase become sizeable is not given by $B_0 = \delta/\mu_B$ but instead by $B^* = \delta^*/\mu_B$ where $\delta^* \gg \delta_0$ is the \textit{typical} (as opposed to minimal) splitting between the Fermi surfaces. For our parameters, we find $\delta^*$ to be of order of $1.5\,\textrm{meV}$ or $B^* \approx 26\,\textrm{T}$ and, thus, negligible in the important field range centered around $B_0$. To check more quantitatively, we use
\begin{equation}
    \phi^s_{\text{Berry}}(B) = \frac{2\pi}{N_c}  \left( 1 - \frac{g \mu_B B/2}{\sqrt{(\delta^*)^2 + (g\mu_B B/2)^2}} \right)
\end{equation}
in \equref{PhasesAccumulated} which is the Berry phase contribution assuming that the splitting of the bands is the same for all momenta on the Fermi surface and given by $\delta^*$. Even for $\delta^*$ significantly smaller than the estimate above, we did not find any noticeable impact of the field-dependence of the Berry phase on the results presented in this work.

\begin{figure}[t]
   \centering
    \includegraphics[width=\linewidth]{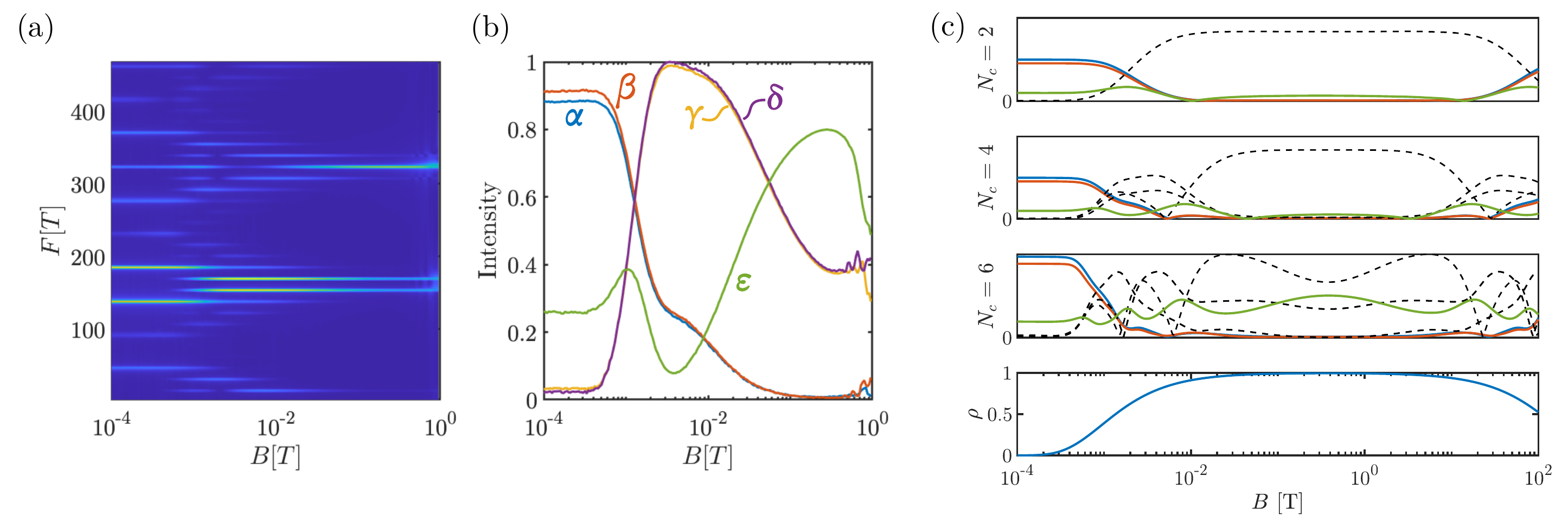}
    \caption{(a) Quantum oscillations frequencies for $N_c=3$ as a function of magnetic field $B$ obtained by Fourier transform of \equref{FunctionalFormToFourierTrafo} in a finite range of inverse magnetic field centered around $1/B$. We use the parameters stated in the text. The intensity of the Fourier transform (normalized to $1$ at the maximum) at the frequencies associated with the five trajectories $\alpha, \beta, \gamma,\delta,\epsilon$ depicted in \figref{fig:QuantumOscillations}(b) are shown in (b). In the upper three panels of (c), we show the intensities in arbitrary units [as in \figref{fig:QuantumOscillations}(c)] at the frequencies $F_1$ and $F_2$ (blue and red) associated with the inner and outer Fermi surface, at the M\"obius frequency $F_1 + F_2$ (green), and at all other expected resonance frequencies $[j F_1 + (N_c-j) F_2]/N_c$, $j=1,2,\dots N_c-1$ (black, dashed) for $N_c=2,3,6$. We use the same parameters as in \figref{fig:QuantumOscillations} and plot the transition probability (fourth panel) as reference.}
    \label{fig:fullquantumosc}
\end{figure} 

\vspace{1em}\noindent\textbf{Computation of quantum oscillations.} To quantify the intensity of the different quantum oscillation frequencies in $1/B$, we use a phenomenological approach similar to \cite{PhysRevB.100.081405}. To construct a signal $I(1/B)$ (mimicking the magnetization's or resistivity's oscillatory behavior with magnetic field) that is peaked whenever (\ref{QuantizationCondition}) is obeyed, we use
\begin{equation}
    \mathcal{S}(1/B) = \frac{\eta}{|\det\left(U(B)-\mathbbm{1}\right)|^2 + \eta^2 }, \label{FunctionalFormToFourierTrafo}
\end{equation}
where $\eta$ is a suitably chosen regularization of the peaks. The quantum oscillation frequency spectra are obtained by Fourier transforming $I(1/B)$ in a finite range $[1/B - \Delta_{B^{-1}},1/B + \Delta_{B^{-1}}]$ of inverse magnetic fields. Using the values for $B_0$ and $\alpha_0$ derived above, taking $\delta^* = 10\,\textrm{T}$ and assuming for concreteness that the area enclosed by the inner and outer Fermi surfaces are $30\,\%$ and $40\,\%$ of the area of the MBZ, we find the Fourier spectrum (taking $\Delta_{B^{-1}} = 1\,\textrm{T}^{-1}$) shown in \figref{fig:fullquantumosc}(a,b). Note that the frequency associated with the M\"obius trajectory is dominant in a wide range of magnetic fields. This includes the largest field we can reach here ($\propto 1/\Delta_{B^{-1}}$). In particular, we cannot present a well-defined spectrum in the regime $B \gtrsim 10\,\textrm{T}$ where the Zeeman effect dominates the transition probability since there are too few oscillations in $\mathcal{S}(1/B)$ in this regime. Although this regime is not of primary interest to our work, we present, for completeness, spectra for it as well in \figref{fig:QuantumOscillations} of the main text [and in \figref{fig:fullquantumosc}(c)]. To be able to do this, we took the limit $A_s \rightarrow \infty$ at fixed $A_+/A_-$ in \figref{fig:QuantumOscillations} where one can keep the value of $B$ fixed at $\bar{B}$ in $\rho(B)$ and $\phi^s_{\text{Berry}}(B)$ when Fourier transforming $\mathcal{S}(1/B)$ in a range centered around $1/\bar{B}$; this provides a well defined quantum oscillation spectrum for any magnetic field. Furthermore, comparison of \figref{fig:QuantumOscillations} and \figref{fig:fullquantumosc}(a-b) shows that the two approaches agree well in the relevant regime $B < 1/\Delta_{B^{-1}}$.

\vspace{1em}\noindent\textbf{Other crystalline point groups.} The fact that the M\"obius trajectory dominates in the large-(orbital)-field limit is a rather unique situation in a 2D crystal; it becomes possible since small-twist-angle graphene moir\'e systems exhibit well defined valley quantum numbers and, within a single valley, both time-reversal, $\Theta$, and out-of-plane two-fold rotation, $C_{2z}$, symmetry are broken. Either one of these two symmetries would guarantee an even number of avoided crossings and the M\"obius trajectory cannot be the dominant one for large orbital magnetic fields where $\rho\rightarrow 1$. In fact, for the same parameters as in \figref{fig:QuantumOscillations} of the main text, we find that the M\"obius trajectory does not dominate for any field, see \figref{fig:fullquantumosc}(c), for the other non-trivial rotational symmetries, $C_{2z}$ ($N_c=2$), $C_{4z}$ ($N_c=4$), and $C_{6z}$ ($N_c=6$), in a crystalline system.

\section{Superconducting instabilities}

\vspace{1em}\noindent\textbf{Parent superconductivity.}  We refer to the superconducting state of the unperturbed tTLG system as the parent superconducting state. The parent state is taken to have intervalley pairing, which is expected to dominate over intravalley pairing due to time-reversal symmetry, $\Theta_s$, and have either SU(2)$_+\times$SU(2)$_-$, or reduced SU(2), spin symmetry. These assumptions are consistent with previous works \cite{XuBalents2018, You2019, OurClassification}. Here SU(2)$_+\times$SU(2)$_-$ refers to independent spin rotations in each valley. Finally, it is assumed that the parent superconductivity occurs only in the partially filled, spin-degenerate bands. We will denote by $\varepsilon^0_{\eta,s,\bm k}$ and $\psi^0_{\eta,s,\bm k}$ the spin-degenerate bands and corresponding eigenfunctions at the Fermi level of the unperturbed system $(h^{(g)}_{\eta}+h^{(t)}_{\eta})$ with quantum numbers: spin $s$; graphene valley $\eta$; and quasimomenta $\bm k$, restricted to the moir\'e BZ.  

\begin{figure}[t!]
{\includegraphics[width=0.3\textwidth,clip]{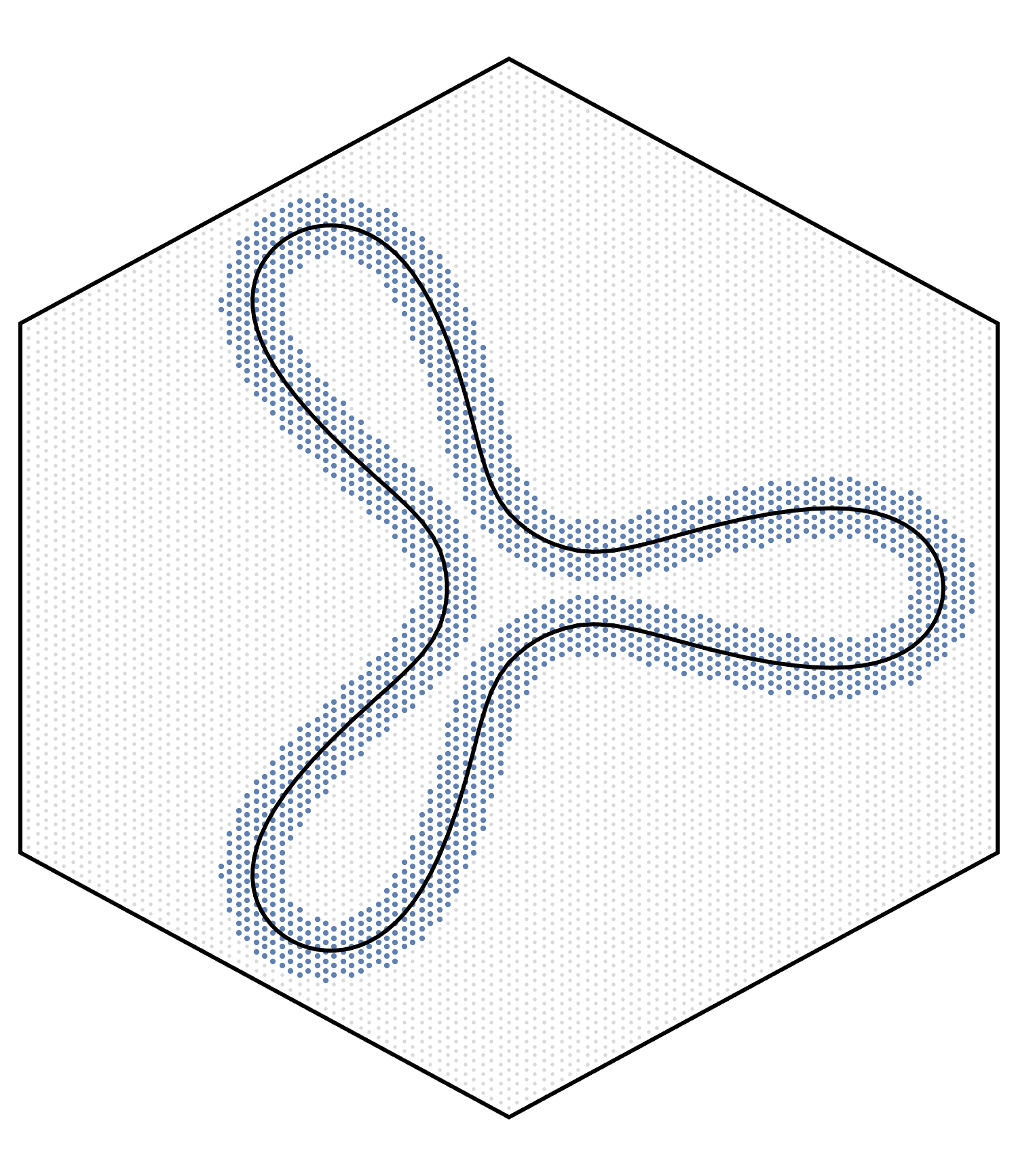}}\hspace{1.25cm}
{\includegraphics[width=0.3\textwidth,clip]{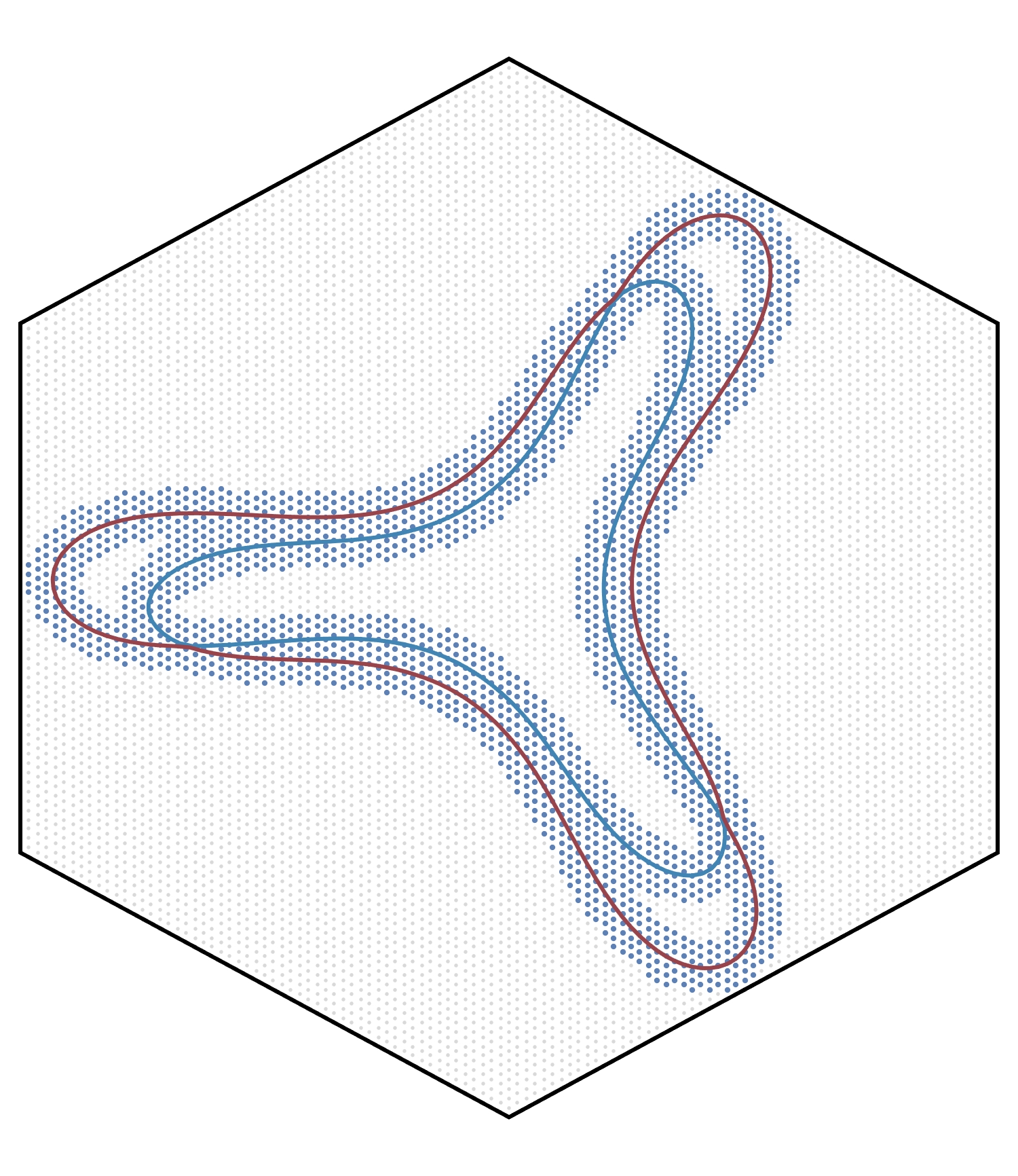}}
\begin{picture}(0,0) 
\put(-340,155){(a)} 
\put(-140,155){(b)} 
\end{picture}
 \caption{The discrete grid realization of the $\Theta_{\bm k;\mu}$-functions of \eqref{GammaAM}. Blue points indicate $\Theta_{\bm k;\mu}=1$, gray points indicate  $\Theta_{\bm k;\mu}=0$. (a) The grid $\Theta_{\bm k;\mu}$-functions used for computations of \figref{fig:SC}; black curve is the Fermi surface of the unperturbed system at $\mu=9.5$ meV (with $\theta=1.50^\circ$).  (b) Grid $\Theta_{\bm k;\mu}$-functions used for \figref{fig:SC}(c); blue and red curves are the spin-split Fermi surfaces of Fig. \ref{fig:Bandstructure}(d) at $\mu=16$ meV (with $\theta=1.75$).}
\label{fig:FSgrid}
\end{figure}
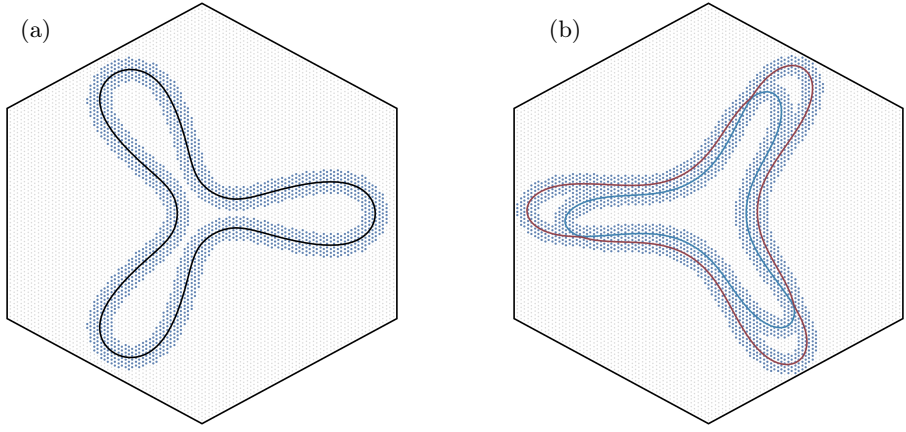

Our ansatz for the pairing interaction, in the Cooper-channel, is 
\begin{align}
\notag {\cal H}_{int}&= \sum_{\bm k_1, \bm k_2} \sum_{s_1,s_2,s_3,s_4}\sum_{\eta=\pm1} (\Gamma_{\bm k_1,\bm k_2})_{s_1,s_2;s_3,s_4} c_{\eta, s_1 \bm k_1}^\dag c_{-\eta,s_3 ,-\bm k_1}^\dag c_{-\eta,s_4,-\bm k_2} c_{\eta,s_2 ,\bm k_2},
\end{align}
The vertex, $(\Gamma_{\bm k_1,\bm k_2})_{s_1,s_2;s_3,s_4}$, is taken to have spin structure,
\begin{align}
    (\Gamma_{\bm k_1,\bm k_2})_{s_1,s_2;s_3,s_4}= \Gamma^{(1)}_{\bm k_1,\bm k_2} (\sigma_0)_{s_1,s_3} (\sigma_0)_{s_4,s_2} + \Gamma^{(2)}_{\bm k_1,\bm k_2} \sum_{i=1}^3(\sigma_i)_{s_1,s_3} (\sigma_i)_{s_4,s_2},
\end{align}
such that $\Gamma^{(1)}_{\bm k_1,\bm k_2}=\Gamma^{(2)}_{\bm k_1,\bm k_2}$ [$\Gamma^{(1)}_{\bm k_1,\bm k_2}\neq\Gamma^{(2)}_{\bm k_1,\bm k_2}$] gives the SU(2)$_+\times$SU(2)$_-$ [SU(2)] limit. The form of the scalar functions $\Gamma^{(1),(2)}_{\bm k_1,\bm k_2}$ depends on the explicit pairing mechanism; for simplicity we assume an Anderson-Morel form, 
\begin{align}
\label{GammaAM}
\Gamma^{(1)}_{\bm k_1,\bm k_2} &= -(\gamma_0+\delta \gamma) \Theta_{\bm k_1;\mu} \Theta_{\bm k_2;\mu},  \quad \Gamma^{(2)}_{\bm k_1,\bm k_2} = -\gamma_0 \Theta_{\bm k_1;\mu} \Theta_{\bm k_2;\mu},
\end{align}
where $\Theta_{\bm k;\mu}$ is a step-function such that: $\Theta_{\bm k;\mu}=1$ for $\bm k$ within a radius $\Lambda$ of any Fermi momentum $\bm k_F$, and $\Theta_{\bm k;\mu}=0$ elsewhere. The $\Lambda$ acts much like the Debye cut-off. Representative plots of $\Theta_{\bm k;\mu}$ are shown in \figref{fig:FSgrid}. An attractive interaction requires $\gamma_0>0$. Meanwhile, the small correction $|\delta\gamma|/\gamma_0<1$ distinguishes three cases of the parent superconducting state: (i) $\delta \gamma>0$ favors spin-singlet, (ii) $\delta \gamma<0$ favors spin-triplet, and (iii) for $\delta \gamma=0$, spin-singlet and triplet are degenerate.

\vspace{1em}\noindent\textbf{Perturbed superconductivity.} We consider now the evolution of the superconducting order under combined perturbation of SOC and $D_0$.  In our construction, the perturbations do not influence the vertex $(\Gamma_{\bm k_1,\bm k_2})_{s_1,s_2;s_3,s_4}$, but do still influence the pairing interaction (and hence gap equation) via the perturbed eigenstates of the noninteracting Hamiltonian, $h_\eta\psi_{\eta,n,\bm k} = \varepsilon_{\eta,n,\bm k}\psi_{\eta,n,\bm k}$, with  $h_\eta=h^{(g)}_{\eta}+h^{(t)}_{\eta} + h^{(\text{SOC})}_\eta+h^{(\text{D})}_\eta$. Here the band index $n$ replaces spin, $s$, which is no longer a good quantum number. Finally, we introduce the electron creation operator $\tilde{c}_{\eta,n,\bm k}^\dag$ for the perturbed system, which is related to the unperturbed creation operator via,
\begin{align}
\label{overlap}
    c^\dag_{\eta,s,\bm k}=C^*_{\eta, n, s, \bm k} \tilde{c}^\dag_{\eta,n,\bm k}, \quad C^*_{\eta, n, s, \bm k}\equiv\psi^\dag_{\eta,n,\bm k} \psi^0_{\eta,s,\bm k}.
\end{align}
    
The mean-field Hamiltonian, decoupled into the Cooper channel for intervalley pairing, is 
  \begin{align}
  \label{supp_HBdG}
  \notag &{\cal H}_\eta =\sum_{\bm k, n} \varepsilon_{\eta,n,\bm k} \tilde{c}_{\eta,n,\bm k}^\dag \tilde{c}_{\eta,n,\bm k} + \sum_{\bm k_1, \bm k_2}\sum_{\mu,\nu} (\Gamma^{-1})_{\bm k_1, \bm k_2; \mu,\nu} d_{\mu, \bm k_1,\eta}^\dag d^\pdagger_{\nu, \bm k_2,\eta} + \\
   & \sum_{\bm k}\sum_{n,n'}\sum_{s_1,s_2}\sum_{\mu,\nu}\Big\{\tilde{c}_{ \eta,n,\bm k}^\dag \tilde{c}_{\eta,n',-\bm k}^\dag \left(d_{\mu,\bm k,\eta} (\sigma_\mu)_{s_1,s_2} C^*_{\eta,n,s_1,\bm k} C_{\eta,n',s_2,\bm k}\right)   +  \text{H.c.} \Big\}.
  \end{align}
  Here the intervalley superconducting order parameter $d_{\mu,\bm k, \eta}$ encodes the momentum and spin structure of the Cooper pairs, where $\mu=0$ refers to spin-singlet and $\mu=1,2,3$ refer to the components of the spin-triplet. Due to SOC, the $\mu$-components mix as a function of $\bm k$; the mixing is encoded in the overlap factors, $C_{\eta, n, s, \bm k}$, of Eq. \eqref{overlap}. Finally, we introduced the more compact (adjoint) notation $\Gamma_{\bm k_1, \bm k_2; \mu,\nu}\equiv -(\gamma_0 \delta_{\mu,\nu}  + \delta\gamma \delta_{\mu,0} \delta_{\nu,0}) \Theta_{\bm k_1;\mu} \Theta_{\bm k_2;\mu}$ and arrive at the linearized gap equation, which follows from \eqref{supp_HBdG}, 
\begin{align}
\label{supp_gapeq}
&d_{\mu, \bm k_1,\eta} = -\sum_{\nu, \bm k_2}\Gamma_{\mu,\mu',\bm k_1,\bm k_2} {\cal W}_{\mu'\nu, \bm k_2,\eta} d_{\nu, \bm k_2,\eta},\\
\notag &{\cal W}_{\mu\nu,\bm k,\eta} = \sum_{n_1,n_2} \sum_{ s_1,s_2,s_3,s_4}\frac{\tanh\left(\frac{\varepsilon_{\eta,n_1\bm k}}{2T}\right)+\tanh\left(\frac{\varepsilon_{\eta,n_2\bm k}}{2T}\right)}{2(\varepsilon_{\eta,n_1\bm k}+\varepsilon_{\eta,n_2\bm k})}(\sigma_\mu)_{s_2,s_3} C_{\eta,n_1,s_1,\bm k} C^*_{\eta,n_1 s_2, \bm k} C_{\eta,n_2,s_3, \bm k} C^*_{\eta,n_2,s_4,\bm k} (\sigma_\nu)_{s_4,s_1}.
\end{align}
Note that the gap equation is diagonal in $\eta$ and, hence, in the main text we specialized to $\eta=+1$; the $\eta=-1$ order parameter can be subsequently deduced from the fermion anticommutation relation.
Using the gap equation in \equref{supp_gapeq}, we compute the evolution of the superconducting order parameters, $d_{\mu, \bm k, \eta}$, under applied SOC and displacement field, as well as for the different spin-symmetries of the parent superconducting state---encoded in the $\delta\gamma$ of \eqref{GammaAM}---leading to the results in \figref{fig:SC} of the main text.

\end{appendix}
\end{document}